\documentclass[12pt, onecolumn]{article}
\usepackage{amsmath}
\usepackage{amssymb}
\usepackage{graphicx}
\usepackage{caption2}
\usepackage{amsfonts}
\usepackage{cite}

\newcommand{\be}{\begin{equation}}
\newcommand{\ee}{\end{equation}}
\newcommand{\bea}{\begin{eqnarray}}
\newcommand{\eea}{\end{eqnarray}}
\newcommand{\bef}{\begin{figure}}
\newcommand{\ef}{\end{figure}}
\newcommand{\bt}{\begin{tabular}}
\newcommand{\et}{\end{tabular}}
\newcommand{\bno}{\begin{enumerate}}
\newcommand{\eno}{\end{enumerate}}

\def\3{\ss}

\catcode`\"=\active
\def"{\accent'177}

\begin{document}

\begin{center}

{\bf \Large  Chaotic motion of three-body problem -- an origin of macroscopic  randomness of the universe }

\vspace{0.3cm}

Shijun  Liao $^{a,b,c,d}$

\vspace{0.3cm}

$^a$ Depart. of Mathematics,   $^b$ State Key Lab of Ocean Engineering\\
$^c$ School of Naval Architecture, Ocean and Civil Engineering\\ Shanghai Jiao Tong University, Shanghai 200240, China

$^d$ Nonlinear Analysis and Applied Mathematics Research Group (NAAM) \\  King Abdulaziz University (KAU), Jeddah, Saudi Arabia 

\end{center}

\hspace{-0.5cm}{\bf Abstract}
{\em  The famous three-body problem is investigated by means of a numerical approach with negligible numerical noises in a long enough time interval,  namely the Clean Numerical Simulation (CNS).    From physical viewpoints,  position of any bodies contains inherent micro-level uncertainty.   The evaluations of such kind of inherent micro-level uncertainty are accurately  simulated  by means of the CNS.   Our  reliable, very accurate CNS results  indicate that the inherent micro-level uncertainty of position of a star/planet might transfer into macroscopic randomness.  Thus, the inherent micro-level uncertainty of a body might be an origin of  macroscopic randomness of the universe.  In addition, from physical viewpoints,  orbits of some three-body systems at large time are  inherently random, and thus it has no physical meanings to talk about the accurate long-term prediction of  the  chaotic  orbits.   Note that such kind of  uncertainty and randomness has nothing to do with the ability of human being.    All of these  might enrich our knowledge and deepen our understandings about not only the three-body problem but also chaos.  }

\hspace{-0.5cm}{\bf Key Words}   Three-body problem,  chaos,   multiple precision,   Taylor expansion,   micro-level uncertainty

\section{Introduction}

It is well-known that the microscopic phenomena are inherent random, while many macroscopic phenomena such as moving stars and planets in the universe looks random as well.   What is the origin of the macroscopic randomness?  Are there any relationships between the microscopic uncertainty and the macroscopic randomness?   In this article,  using  chaotic motion  of  the famous three-body problem as an example,  we illustrate that the micro-level uncertainty of position of stars/planets might be one origin of the macroscopic randomness of the universe.  

It is a common knowledge that some ``deterministic''  dynamic systems have chaotic property: their numerical simulations have sensitive dependence on initial conditions (SDIC), i.e. the so-called butterfly-effect, so that long-term accurate prediction is impossible \cite{Lorenz1963,  Lorenz1993, Lorenz2006}.   Since truncation and round-off errors are unavailable  for all numerical simulation techniques, nearly all numerical results given by traditional methods  based on double precision are not ``clean'':  they are something mixd with the so-called  ``numerical noises''.   Due to the SDIC,  truncation and round-off errors enlarge exponentially  so that it is very hard to gain reliable chaotic results in a long time interval.   As pointed  out by Lorenz \cite{Lorenz2006} in 2006,   different traditional numerical schemes (based on 16 or 32 digit precision) may lead to not only the uncertainty in prediction but also  fundamentally different regimes of solution.  

In order to gain reliable chaotic results in a long enough time interval, Liao \cite{Liao2009} developed a numerical technique with negligible numerical noises,  called  the  ``Clean Numerical Simulation'' (CNS).  Using the computer algebra system Mathematica with the 400th-order Taylor expansion and data in 480-digit precision, Liao \cite{Liao2009} obtained, for the first time,  the reliable numerical results of chaotic solution of Lorenz equation  in a long time interval $0 \leq   t  \leq  1000$  Lorenz time unit (LTU).    Liao's ``clean'' chaotic solution of Lorenz equation  was  confirmed by Wang et al. \cite{Wang2011},  who  employed   the  parallel computation  and  the multiple precision (MP) library  to gain reliable chaotic solution up to 2500 LTU by means of the CNS approach with 1000th-order Taylor expansion and data in 2100-digit precision,  and their result agrees well with Liao's one \cite{Liao2009} in $0 \leq  t \leq  1000$ LTU.   This confirms the validity of the CNS approach.  

It was found by Liao \cite{Liao2009} that, to gain a reliable ``clean'' chaotic solution of Lorenz equation in the interval  $0 \leq  t  \leq T_c$,   the initial conditions must be at least in the accuracy of $10^{-2 T_c/5}$.  For example,   in the case of $ T_c = 1000$  LTU,  the initial condition must be in 400-digit precision at least.   It should be emphasized that,  the 400-digit precision,  which is ``mathematically'' necessary for the initial condition and all  data at each time-step,  is so high that even the statistical  fluctuation of velocity and temperature becomes a very  important physical factor and therefore cannot be neglected.  However,  from the physical viewpoints,   Lorenz equation  (as a macroscopical model for climate prediction on Earth)  completely neglects the influence of the statistic fluctuation of  velocity and temperature  about   the climate.   Therefore, as pointed out by Liao \cite{Liao2009},  this  leads to  the so-called ``precision paradox of chaos''.    
  
How to avoid such kind of paradox?    Traditionally,  it  is  believed that  Lorenz equation is a ``deterministic'' system, say,  its initial condition and all physical parameters are completely certain, i.e.  ``absolutely  accurate''.   However,  such kind of ``absolutely accurate'' variables only exist in mathematics,  which have no physical meanings in practice, as mentioned below.   For example,   velocity and temperature of fluid  are concepts defined by statistics.   It is well-known that any statistical variables contain statistic fluctuation.    So, strictly speaking,  velocity and temperature of fluid are not ``absolutely accurate'' in physics.   Using Lorenz equation as an example, Liao \cite{Liao2012}   pointed out that its initial conditions  have fluctuations in the micro-level $10^{-30}$  so  that  Lorenz equation is {\em not} deterministic, from the {\em physical} viewpoint.   Although $10^{-30}$ is  much smaller  than  truncation and round-off errors of traditional numerical methods based on double precision,  it is much larger than $10^{-400}$ that can be used in the CNS approach.   
Thus,  by means of the CNS,  Liao \cite{Liao2012}  accurately  simulated  the evaluation of the micro-level uncertainty of initial condition of Lorenz equation, and found that  the micro-level uncertainty transfers into the observable  randomness.  Therefore, chaos might be a bridge between the micro-level uncertainty and the macroscopic randomness, as  pointed out by Liao \cite{Liao2012}.   Currently, Liao \cite{Liao2012-CFS}  employed the CNS  to the chaotic Hamiltonian H\'{e}non-Heiles system for motion of stars orbiting in a plane about the galactic center, and confirmed that, due to the SDIC, the inherent micro-level uncertainty of position of stars indeed evaluates into the macroscopic randomness.           

However, Lorenz equation is a  greatly  simplified model of Navier-Stokes equation for flows of fluid.   Besides,  unlike Hamiltonian H\'{e}non-Heiles system for motion of stars orbiting in a plane,  orbits of stars are three dimensional in practice.    Thus, in order to further confirm the above conclusion,   it is necessary to investigate some more accurate physical models, such as the famous three-body problem \cite{Henon1964, Diacu1996, Valtonen2005}  governed by the Newtonian gravitation law.   In fact,  non-periodic  results  were  first  found  by Poincar\'{e} \cite{Poincare1890} for three-body problem.    In this paper, using the three-body problem as a better physical model,  we employ the CNS to confirm the conclusion:  the {\em inherent} micro-level uncertainty of position of a star/planet might transfer into the observable, macroscopic randomness of its orbit so that the {\em inherent} micro-level uncertainty of a star/planet might be an origin of the macroscopic randomness of the universe.

\section{Approach of Clean Numerical Simulation}

Let us consider the famous  three-body problem, say, the motion of three celestial bodies under their mutual gravitational attraction.    Let $x_1, x_2, x_3$ denote the three orthogonal axises.  The  position  vector  of  the $i$ body is expressed by
${\bf r}_i = (x_{1,i},x_{2,i},x_{3,i})$.   Let $T$ and $L$ denote the characteristic time and length  scales,   and $m_i$  the mass of the $i$th  body,  respectively.    Using Newtonian gravitation law,  the motion of the three bodies are governed by the corresponding non-dimensional equations
\begin{equation}
\ddot{x}_{k,i} = \sum_{j=1,  j\neq i}^{3}  \rho_j \frac{(x_{k,j}-x_{k,i})}{R_{i,j}^{3}},\;\;\; k = 1,2,3,  \label{geq:x[k,j]}
\end{equation}  
where
\begin{equation}
R_{i,j} = \left[ \sum_{k=1}^{3} (x_{k,j}-x_{k,i})^2 \right]^{1/2}   \label{def:R[i,j]}
\end{equation}
and 
\begin{equation}
 \rho_i = \frac{m_i}{m_1}, \;\; i = 1,2,3 
\end{equation}
denotes the ratio of the mass.  

In the frame of the CNS,  we use  the $M$-order Taylor expansion 
\begin{equation}
x_{k,i}(t) \approx \sum_{m=0}^{M} \alpha_m^{k,i} \; (t-t_0)^m \label{series:x[k,i]}
\end{equation}
to accurately calculate the orbits of the three bodies, where the coefficient  $\alpha_m^{k,i}$ is only dependent upon the time $t_0$.     Note that the position  $x_{k,i}(t)$ and velocity $\dot{x}_{k,i}(t)$ at $t=t_0$ are known, i.e.
\begin{equation}
\alpha_0^{k,i}  = x_{k,i}(t_0),  \;\;  \alpha_1^{k,i}  = \dot{x}_{k,i}(t_0). 
\end{equation}
 The recursion formula of $\alpha_m^{k,i}$ for $m\geq 2$ is derived from (\ref{geq:x[k,j]}), as described  below. 

Write  $1/R_{i,j}^3$ in the Taylor expansion  
\begin{equation}
f_{i,j} = \frac{1}{R_{i,j}^3} \approx \sum_{m=0}^{M} \beta_m^{i,j} \; (t-t_0)^m  \label{series:1/R[i,j]}
\end{equation}
with the symmetry property $\beta_m^{i,j} =\beta_m^{j,i}$, where $\beta_m^{i,j} $ is determined later.      
Substituting (\ref{series:x[k,i]}) and (\ref{series:1/R[i,j]}) into (\ref{geq:x[k,j]}) and comparing the like-power of $(t-t_0)$, we have the recursion  formula
\begin{equation}
\alpha_{m+2}^{k,i} = \frac{1}{(m+1)(m+2)} \sum_{j=1, j\neq i}^{3} \rho_j \sum_{n=0}^{m} \left( \alpha_n^{k,j}-\alpha_n^{k,i}\right) \; \beta_{m-n}^{i,j}, \;\; m \geq 0.  
\end{equation}
Thus, the positions and velocities of the three bodies  at the next time-step $t_0+\Delta t$ read 
\begin{eqnarray}
x_{k,i}(t_0+\Delta t) &\approx & \sum_{m=0}^{M} \alpha_m^{k,i} \; (\Delta t)^m  \label{result:position} ,\\
\dot{x}_{k,i}(t_0+\Delta t) &\approx & \sum_{m=1}^{M} m \; \alpha_m^{k,i} \; (\Delta t)^{m-1} \label{result:x[k,i]} . \label{result:velocity} 
\end{eqnarray}

Write 
\begin{equation}
 S_{i,j} = R^6_{i,j} \approx \sum_{m=0}^{M} \gamma_m^{i,j} \; (t-t_0)^m, \;\;  f_{i,j}^2 = \sum_{m=0}^{M} \sigma_{m}^{i,j} (t-t_0)^m,  \label{series:S[i,j]}
 \end{equation}
 with the symmetry property $\gamma_m^{i,j}=\gamma_m^{j,i}$ and $\sigma_m^{i,j}=\sigma^{j,i}_m$.   Substituting (\ref{def:R[i,j]}), (\ref{series:x[k,i]}) and (\ref{series:1/R[i,j]}) into the above definitions and comparing the like-power of $t-t_0$,  we have 
\begin{eqnarray}
\gamma^{i,j}_m &=& \sum_{n=0}^{m} \mu^{i,j}_{m-n} \sum_{k=0}^{n}\mu_k^{i,j} \mu_{n-k}^{i,j},    \\
\sigma_m^{i,j} &=& \sum_{n=0}^m \beta^{i,j}_n \; \beta^{i,j}_{m-n},   
\end{eqnarray}
with
\begin{equation}
\mu^{i,j}_m = \sum_{k=1}^{3} \sum_{n=0}^{m}\left( \alpha^{k,j}_{n}-\alpha^{k,i}_{n}\right)\left( \alpha^{k,j}_{m-n}-\alpha^{k,i}_{m-n}\right), \;\; i\neq j, \; m\geq 1, 
\end{equation}
and the symmetry   $\mu_m^{i,j}=\mu_m^{j,i}$. 
Using the definition (\ref{series:1/R[i,j]}),  we have 
\[   S_{i,j} f_{i,j}^2 = 1.   \]   
Substituting  (\ref{series:S[i,j]}) into the above equation  and comparing the like-power of $(t-t_0)$, we have 
\[    \sum_{n=0}^{m} \gamma_n^{i,j} \; \sigma_{m-n}^{i,j} = 0,  \;\; m\geq 1, \]
which gives the recursion formula
\begin{equation}
\beta_m^{i,j} = -\frac{1}{2\beta_0^{i,j} \gamma_0^{i,j}} \left\{\sum_{n=1}^{m} \gamma_n^{i,j}\sigma_{m-n}^{i,j} +\gamma_0^{i,j}\sum_{k=1}^{m-1}\beta_k^{i,j}\beta_{m-k}^{i,j}\right\}, \;\; m\geq 1, \; i\neq j.
\end{equation}
 In addition,  it is straightforward that
\begin{equation}
\beta_0^{i,j} = \frac{1}{R_{i,j}^3}, \;\;\;  \mu_0^{i,j} = R_{i,j}^2, \;\;\; \sigma_0^{i,j} = \left( \beta_0^{i,j} \right)^2,  \;\;\;  \gamma^{i,j}_0 =  \left( \mu_0^{i,j}\right)^3,\;\;\; \mbox{at $t=t_0$}.    
\end{equation}

It is a common knowledge that numerical methods always contain truncation  and round-off errors.    To decrease the round-off error,  we express  the positions,  velocities, physical parameters and {\em all} related data  in $N$-digit precision, where $N$ is a large enough positive integer.   Obviously, the larger the value of $N$, the smaller the round-off error.  Besides,  
 the higher the order $M$ of Taylor expansion (\ref{series:x[k,i]}),  the smaller the truncation error.    Therefore,  if the order $M$ of Taylor expansion (\ref{series:x[k,i]}) is high enough and all data are expressed in  accuracy of long enough digits,   truncation and round-off errors of the above-mentioned CNS approach (with reasonable time step $\Delta t$) can be so small that numerical noises are negligible in a given (long enough) time interval, say,  we can gain ``clean'', reliable chaotic numerical results with certainty in a given time interval.    In this way,  the orbits of the three bodies can be calculated, accurately and correctly,  step by step.    
 
In this article, the computer algebra system Mathematica is employed.   By means of the Mathematica,  it is rather convenient to express all datas in 300-digit  precision, i.e. $N=300$.   In this way, the round-off error is so small that it is almost negligible in the given time interval ($0\leq t\leq 1000$).  And the accuracy of the CNS  results  increases as the order $M$  of Taylor expansion (\ref{series:x[k,i]}) enlarges, as shown in the next section.        

\section{A special example}

Without loss of generality,  let us consider the motion of three bodies  with the initial positions
\begin{equation}
 {\bf r}_1 = (\delta ,0,-1),  {\bf r}_2 = (0,0,0),  {\bf r}_3 = - ({\bf r}_1  + {\bf r}_2),  \label{ic:r}
\end{equation}
and the initial velocities 
\begin{equation}
\dot{\bf r}_1 = (0,-1,0), \dot{\bf r}_2 = (1,1,0), \dot{\bf r}_3 = -(\dot{\bf r}_1 + \dot{\bf r}_2),  \label{ic:velocity}
\end{equation}
where $\delta $ is a constant.  Note that $\delta$ is the only one unknown parameter in the initial condition.   For simplicity,  we first only consider the  three different cases:  $\delta  = 0$,   $\delta =  +10^{-60}$ and $\delta =  - 10^{-60}$.   Note that the initial velocities are the same in the three cases.     {\em Mathematically},  the  three initial positions have the tiny difference in the level of $10^{-60}$,  which however leads to huge difference of orbits of the three bodies at $t=1000$, as shown below.    For the sake of simplicity,  let us consider the case of the three bodies with equal masses, i.e. $\rho_j = 1$ ($j=1,2,3$).     We are interested in the orbits of the three bodies in the time interval $0\leq t \leq 1000$. 

Note that the initial conditions satisfy 
\[    \sum_{j=1}^3 \dot{\bf r}_j(0)  = \sum_{j=1}^3 {\bf r}_j(0) = 0.  \]
Thus, due to the momentum conversation, we have  
\begin{equation}
\sum_{j=1}^3 \dot{\bf r}_j(t)  = \sum_{j=1}^3 {\bf r}_j(t) = 0, \hspace{1.0cm} t \geq 0  \label{conversation:r}
\end{equation}
in general.

All data are expressed in 300-digit precision, i.e. $N=300$.  Thus, the round-off error is  almost negligible.   
In addition, the higher the order $M$ of Taylor expansion (\ref{series:x[k,i]}),  the smaller the truncation error, i.e.  the more accurate the results at $t=1000$.     Assume that,  at $t=1000$,  we have   the result $x_{1,1} = {\bf 1.81510} 12345$    by means of the $M_1$-order Taylor expansion and the result $x_{1,1} = {\bf 1.81510} 47535$  by means of the $M_2$-order Taylor expansion,  respectively,  where $M_2 > M_1$.  Then,  the result $x_{1,1}$  by means of the lower-order ($M_1$) Taylor expansion  is said to be in the accuracy of 5 significance digit,  expressed by $n_s = 5$.       For more details about the CNS, please refer to Liao \cite{Liao2012-CFS}.  

\begin{figure}
\centering
\includegraphics[scale=0.4]{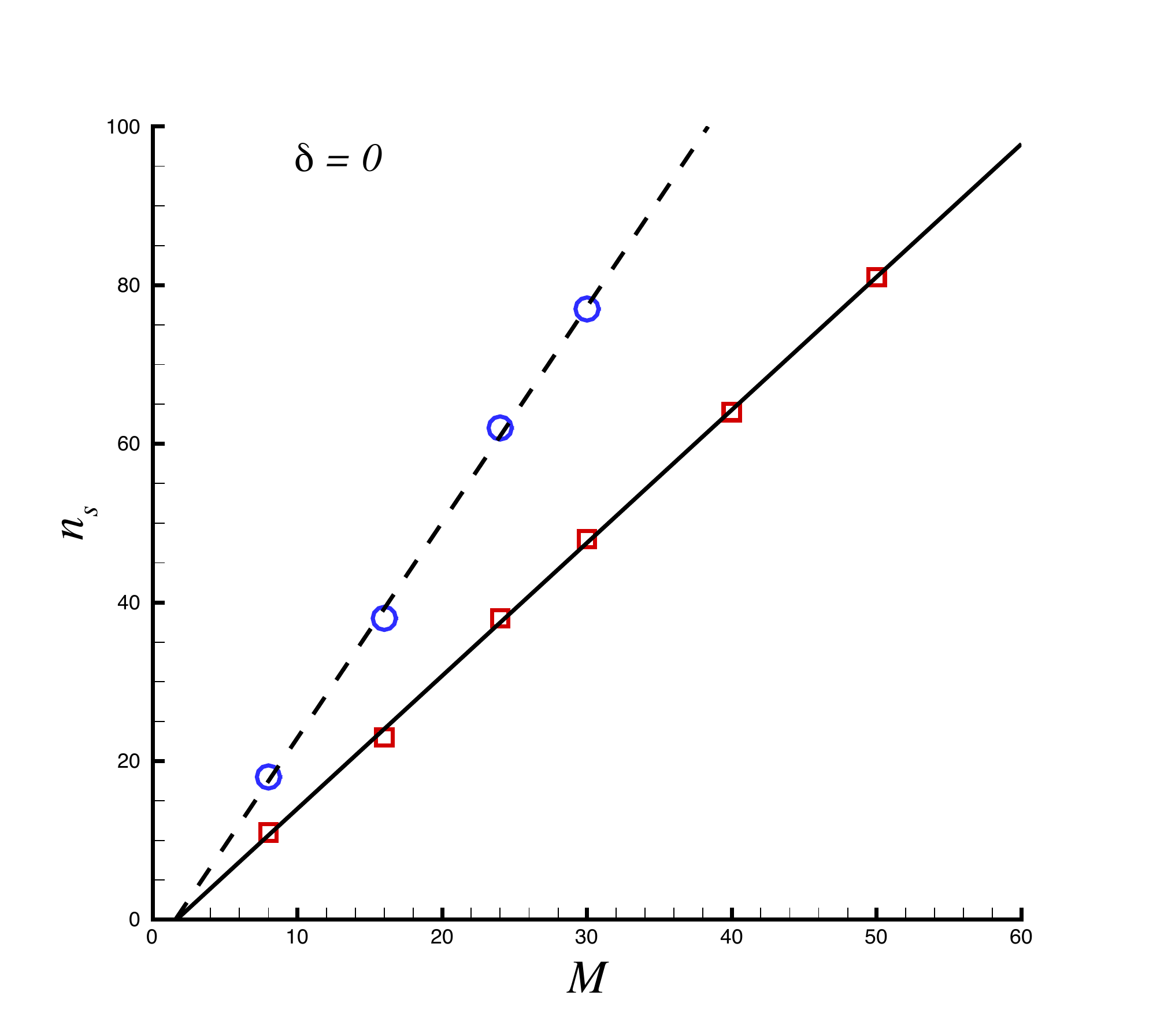}
\caption{ Accuracy (expressed by $n_s$, number of significance digits) of the CNS results at $t=1000$  in the case of $\delta = 0$ by means of $N=300$, the different time-step $\Delta t$ and the different order ($M$) of Taylor expansion.  Square: $\Delta t = 10^{-2}$; Circle: $\Delta t = 10^{-3}$.  Solid line: $n_s \approx 1.6762M - 2.7662$; dashed line: $n_s \approx  2.7182 M - 4.2546$.   }
\label{figure:accuracy:delta-0}
\end{figure}

\begin{figure}
\centering
\includegraphics[scale=0.3]{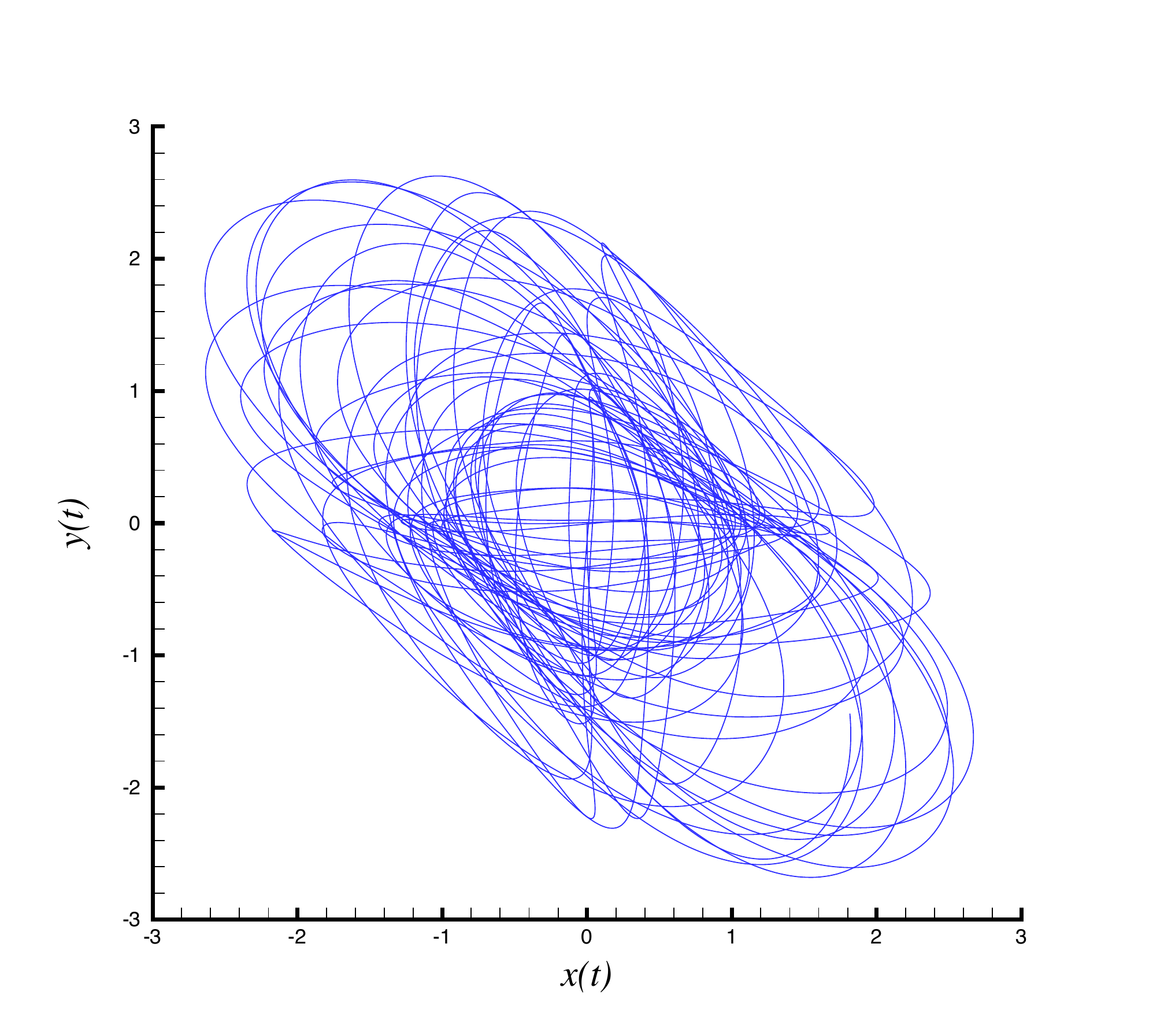}
\includegraphics[scale=0.3]{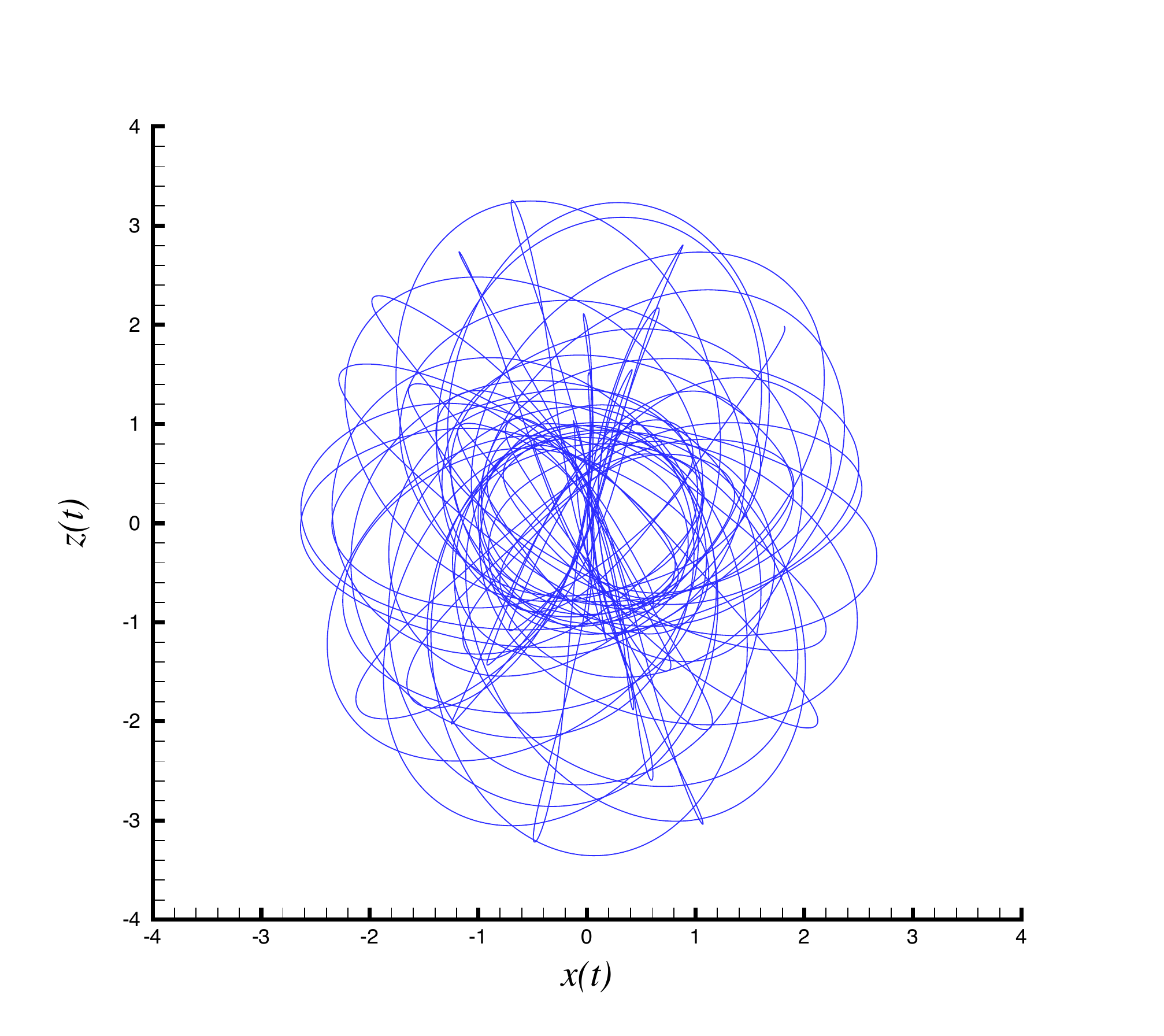}
\caption{ $x-y$ and $x-z$ of Body 1  ($0 \leq t \leq 1000$)  in the case of $\delta = 0$.}
\label{figure:body1-2D}

\centering
\includegraphics[scale=0.3]{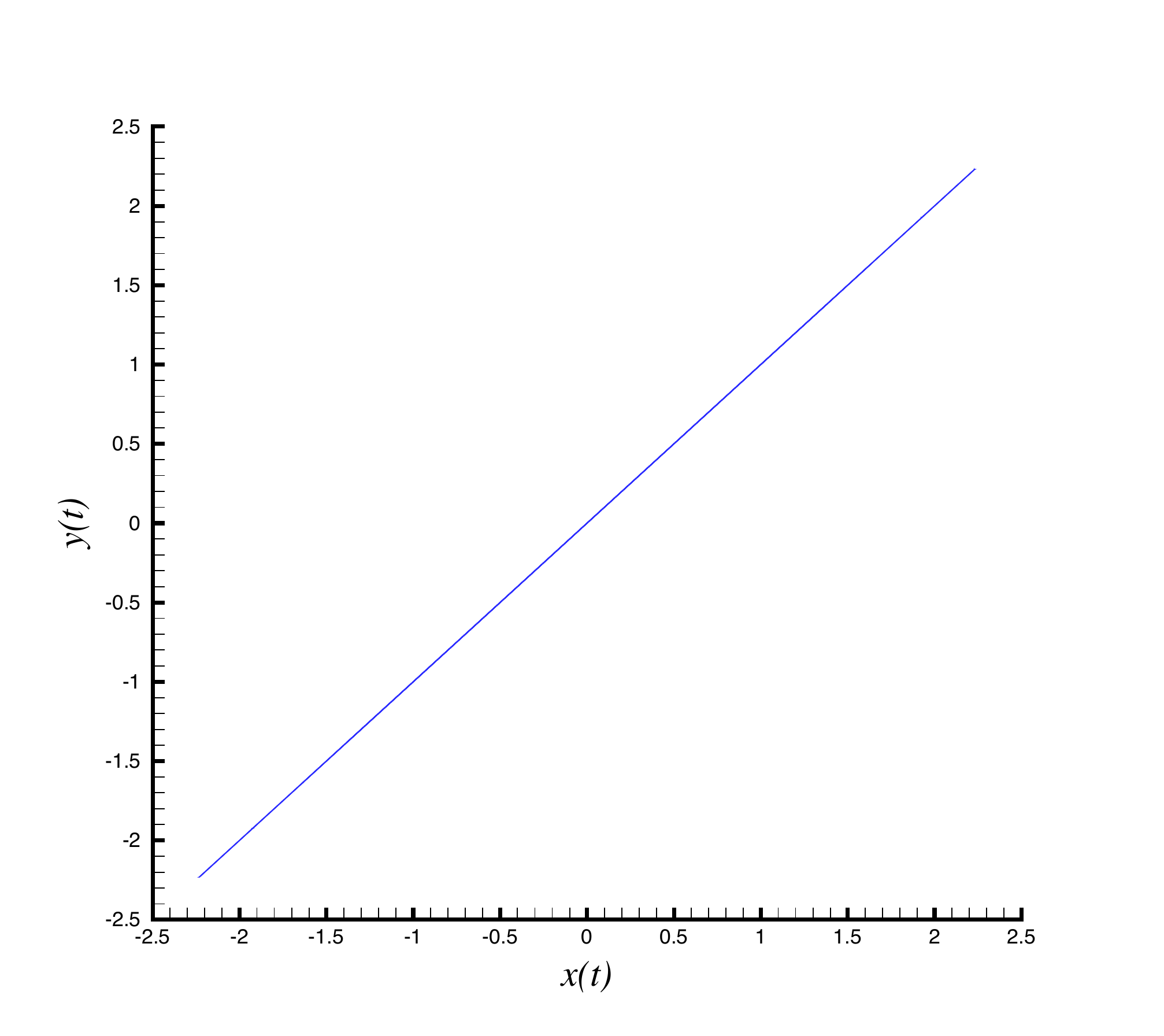}
\includegraphics[scale=0.3]{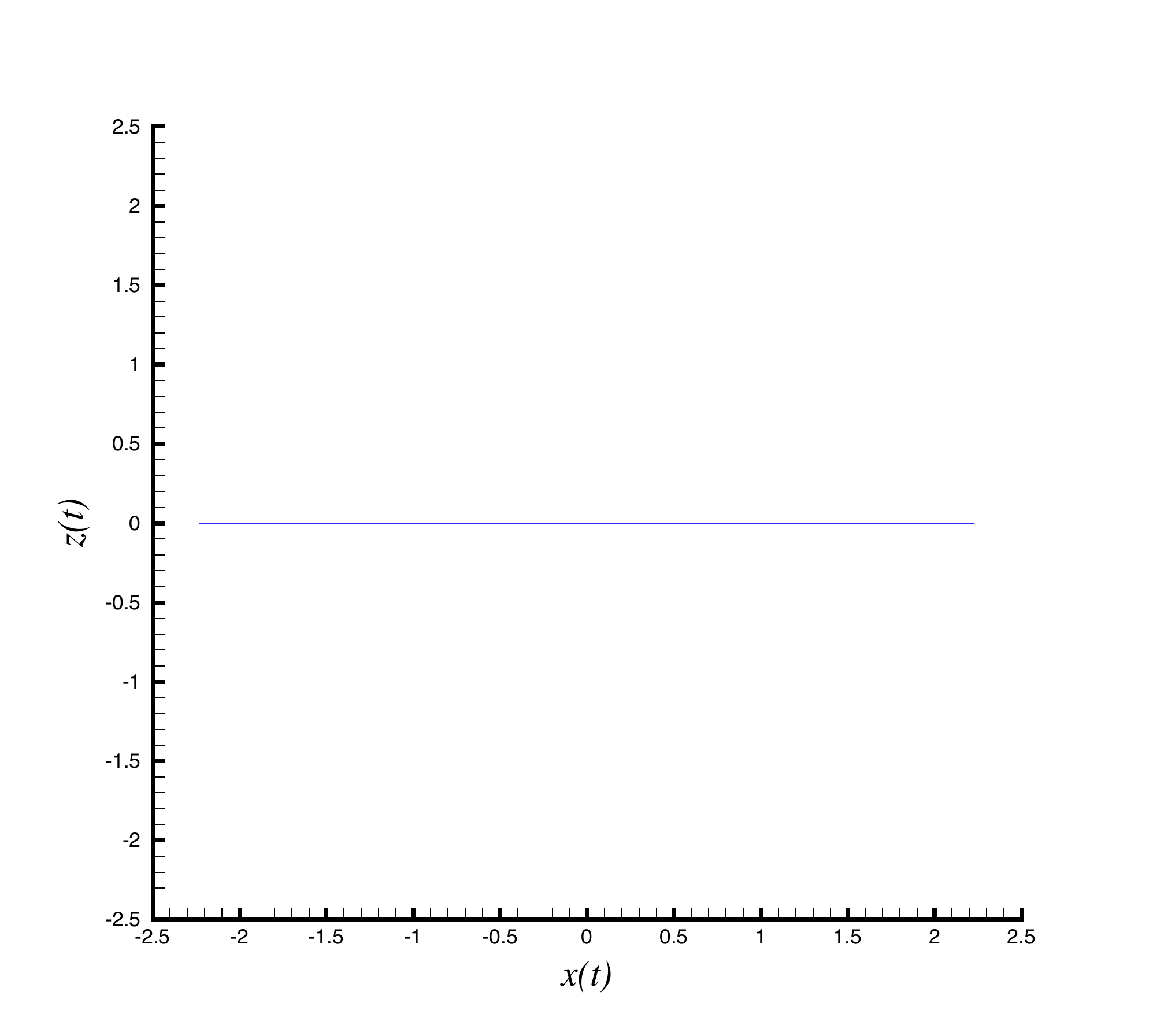}
\caption{ $x-y$ and $x-z$ of Body 2  ($0 \leq t \leq 1000$) in the case of $\delta = 0$. }
\label{figure:body2-2D}

\centering
\includegraphics[scale=0.3]{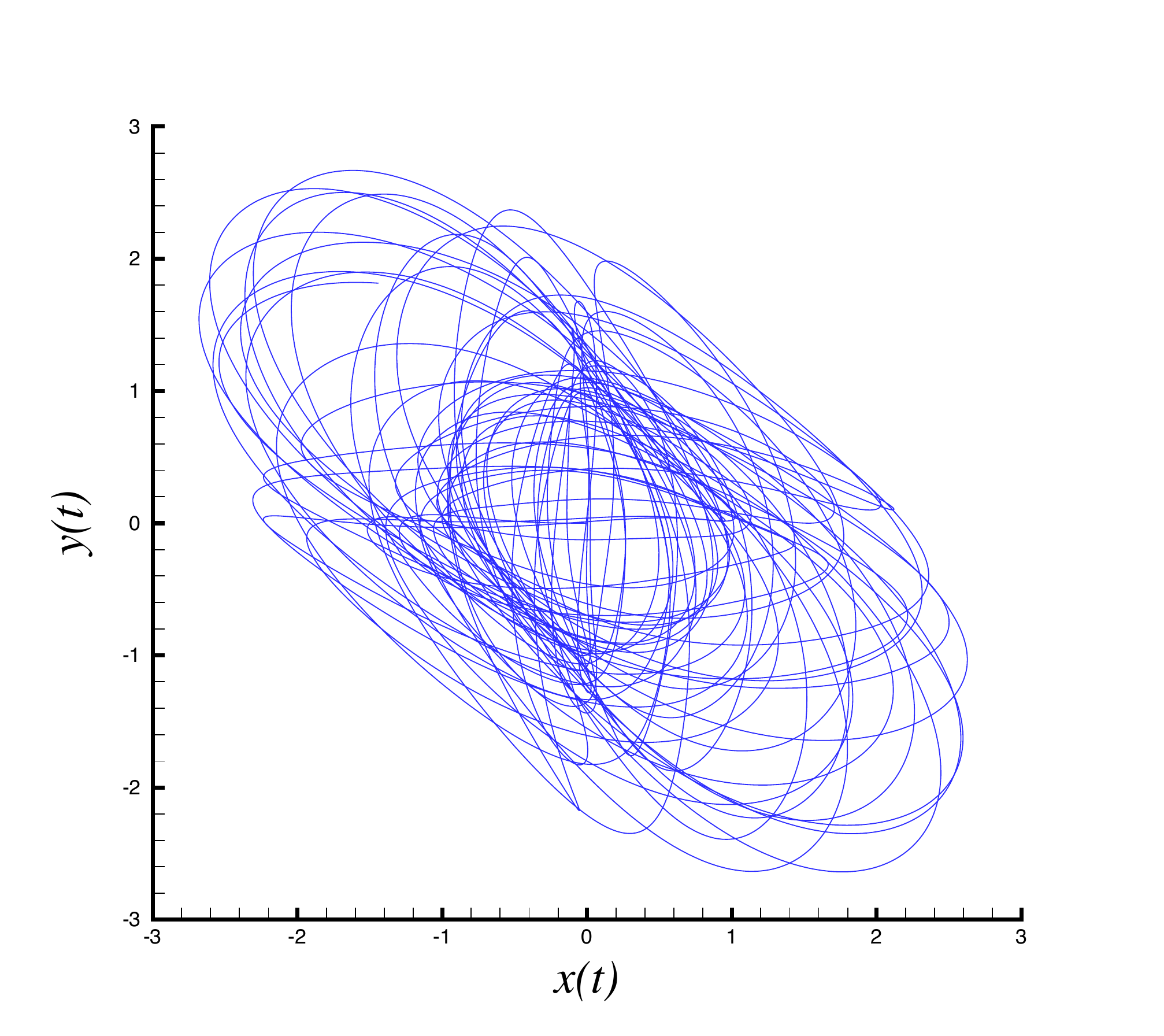}
\includegraphics[scale=0.3]{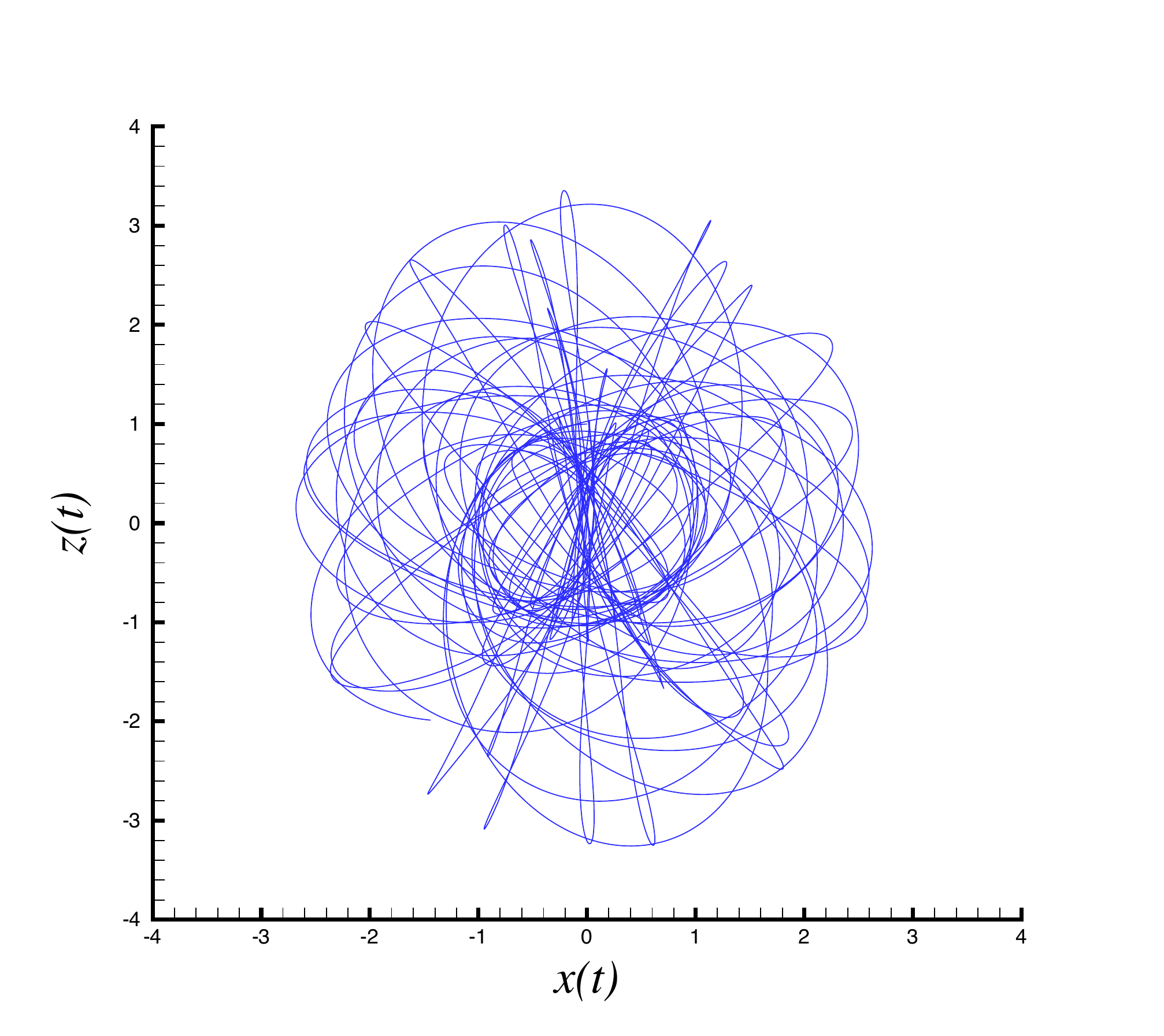}
\caption{ $x-y$ and $x-z$ of Body 3  ($0 \leq t \leq 1000$)  in the case of $\delta = 0$. }
\label{figure:body3-2D}
\end{figure}   

When $\delta = 0$, the corresponding three-body problem has chaotic orbits with the Lyapunov exponent  $\lambda = 0.1681$, as pointed by Sprott \cite{Sprott2010} (see  Figure 6.15 on page 137).    It is well-known that a chaotic dynamic system has  the sensitivity dependence on initial condition ( SDIC).    Thus, in order to gain reliable   numerical results of the  chaotic orbits  in such a long interval $0\leq t \leq 1000$,  we employ  the  CNS approach using the high-enough order $M$  of Taylor expansion with all data  expressed in 300-digit precision.      

It is found that, when $\delta=0$,  the CNS  results at $t=1000$ by means of $\Delta t = 10^{-2}$, $N = 300$ and $M= 8,16, 24, 30, 40$ and 50  agree each other  in the accuracy of   11,  23,  38,  48,  64  and 81  significance digits, respectively.    Approximately,  $n_s$, the number of significance digits of the positions at $t=1000$,   is linearly  proportional to $M$ (the order of Taylor expansion), say, $n_s \approx 1.6762 M - 2.7662$,  as shown in Fig.~\ref{figure:accuracy:delta-0}.     For example,  the CNS approach using the 50th-order Taylor expansion and data in 300-digit precision (with $\Delta t=10^{-2}$)  provides us  the position of Body 1  at  $t=1000$   in the accuracy of 81 significance digit:  
\begin{eqnarray}
x_{1,1} &=&  
  +1.81510 47535 62951 61721 65940 08845 44006 45690 \nonumber \\ && 03032 05574  32375 90103 
28524 04361 81354 98683 4,  \label{x[1,1]-delta-0}\\
x_{2,1} &= & -1.44063 51440 58286 16113 38350 55406  74890 01231 \nonumber \\ && 29002 84853 
72197 68908 17632 27032 47482 40993 8, \\
x_{3,1} & = & + 1.98700 78875 78629 88109 76776 41498 57789  79168 \nonumber \\ &&  46299 74639 70517 57074.  
11773 08217 34512 84475 9. \label{x[3,1]-delta-0}
\end{eqnarray}
Using the smaller time step $\Delta = 10^{-3}$ and data in 300-digit precision (i.e. $N=300$),  the CNS results given by the 8, 16, 24 and 30th-order Taylor expansion agree in the accuracy of 18, 38, 62 and 77 significance digits, respectively.  Approximately,  $n_s$, the number of significance digits of the positions at $t=1000$,   is linearly  proportional to $M$ (the order of Taylor expansion), say, $n_s \approx 2.7182 M - 4.2546$,  as shown in Fig.~\ref{figure:accuracy:delta-0}.   It should be emphasized that the CNS results  by $\Delta t = 10^{-3}$ and $M=30$ agree (at least)  in the 77 significance digits  with those by $\Delta t = 10^{-2}$ and  $M=50$ in the {\em whole} time  interval $0\leq t \leq 1000$.   In addition, the momentum conservation (\ref{conversation:r}) is satisfied in the level of $10^{-295}$.   All of these  confirm the correction and reliableness  of our  CNS results\footnote{Liao \cite{Liao2012-CFS} proved a convergence-theorem and explained the validity and reasonableness of the CNS by using the mapping $x_{n+1} = \mbox{mod}(2 x_n,1)$.}.   Thus, although the  considered  three-body problem has chaotic orbits,   our numerical results given by the CNS  using  the 50th-order Taylor expansion and accurate data in 300-digit precision (with $\Delta t=10^{-2}$) are reliable in the accuracy of 77 significance digits  in the whole interval $0\leq t \leq 1000$.      

The orbits of the three bodies in the case of $\delta=0$ are as shown in Figs.~\ref{figure:body1-2D} to \ref{figure:body3-2D}.    The orbits of Body~1 and Body~3 are chaotic.  This agrees well with Sprott's conclusion  \cite{Sprott2010} (see  Figure 6.15 on page 137).   However, it is interesting that Body~2  oscillates along a line on the plane $z = 0$.    So, since  $\sum_{j=1}^3 \dot{\bf r}_j$   = $\sum_{j=1}^3{\bf r}_j$   = 0 due to the momentum conservation,  the  chaotic  orbits  of  Body~1 and Body~3 must be  symmetric  about the  regular  orbit  of Body~2.   Thus, although the orbits of Body~1 and Body~3  are  disorderly, the three bodies as a system have an elegant structure with symmetry.

 \begin{figure}
\centering
\includegraphics[scale=0.4]{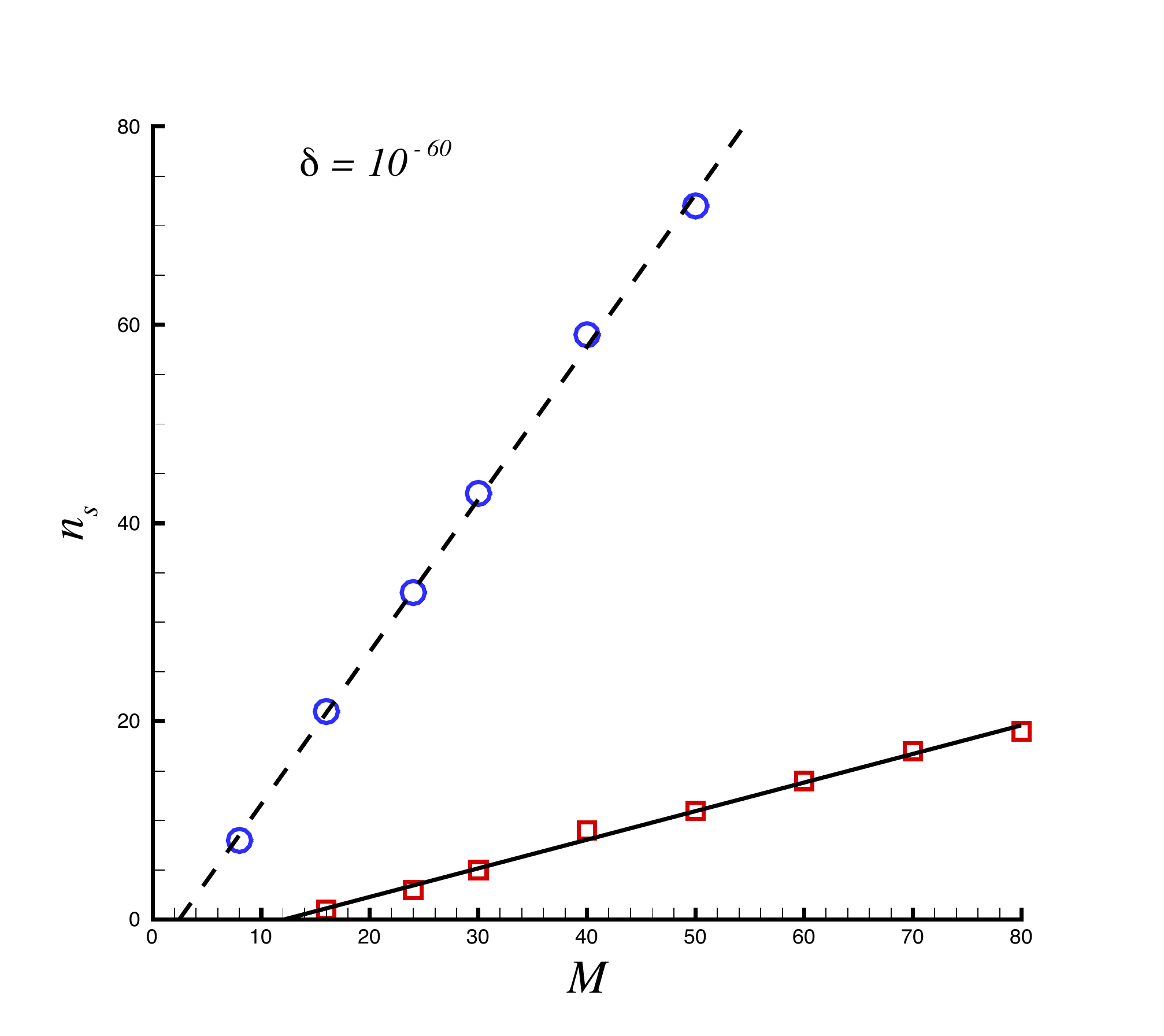}
\caption{ Accuracy (expressed by $n_s$, number of significance digits) of the result at $t=1000$  in the case of $\delta = 10^{-60}$ by means of $N=300$, the different time-step $\Delta t$ and the different order of Taylor expansion ($M$).  Square: $\Delta t = 10^{-2}$; Circle: $\Delta t = 10^{-3}$.  Solid line: $n_s \approx 0.2885M - 3.4684$; dashed line: $n_s \approx 1.5386 M - 3.7472$.   }
\label{figure:accuracy:delta-60}
\end{figure}

\begin{figure}
\centering
\includegraphics[scale=0.3]{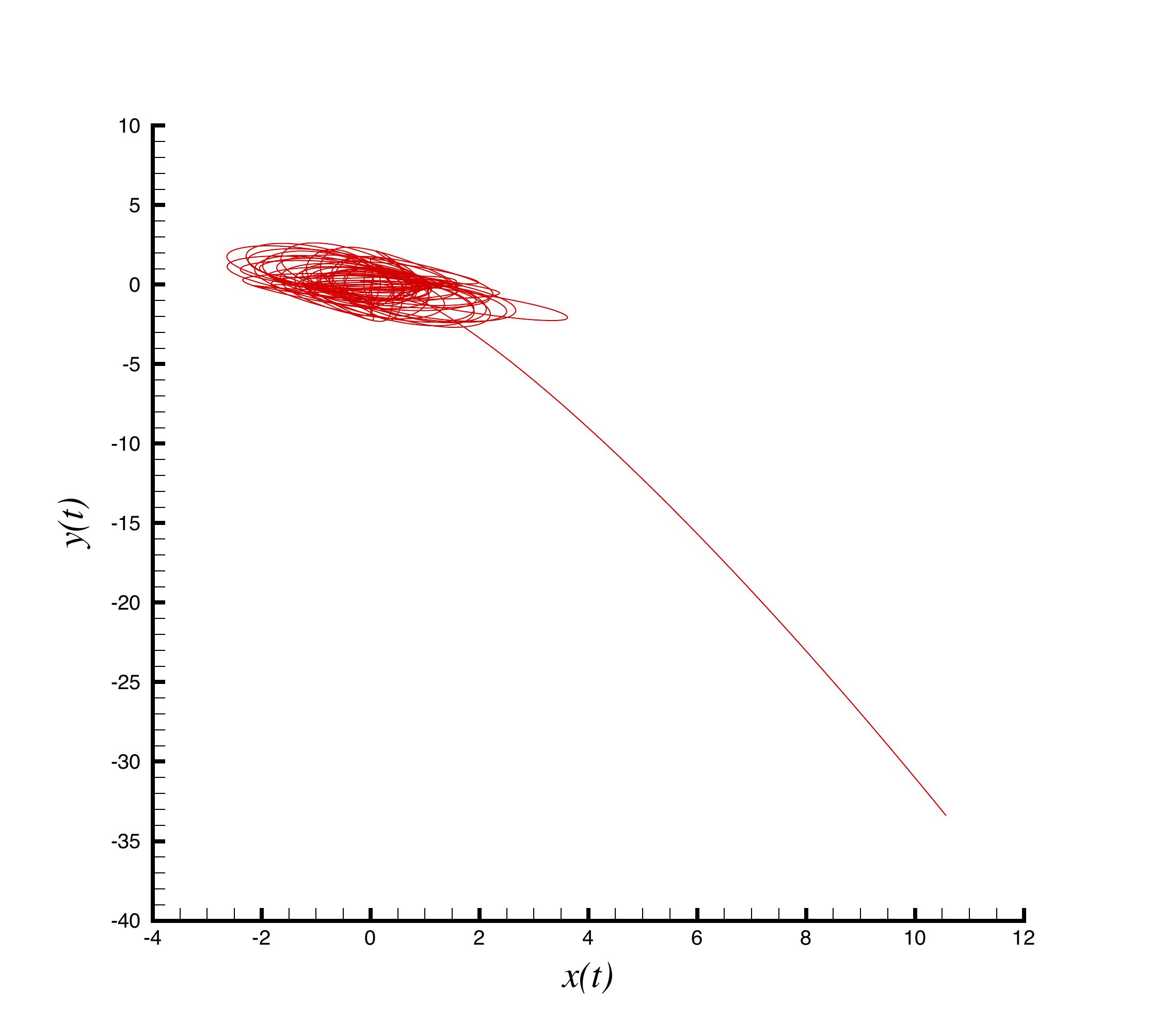}
\includegraphics[scale=0.3]{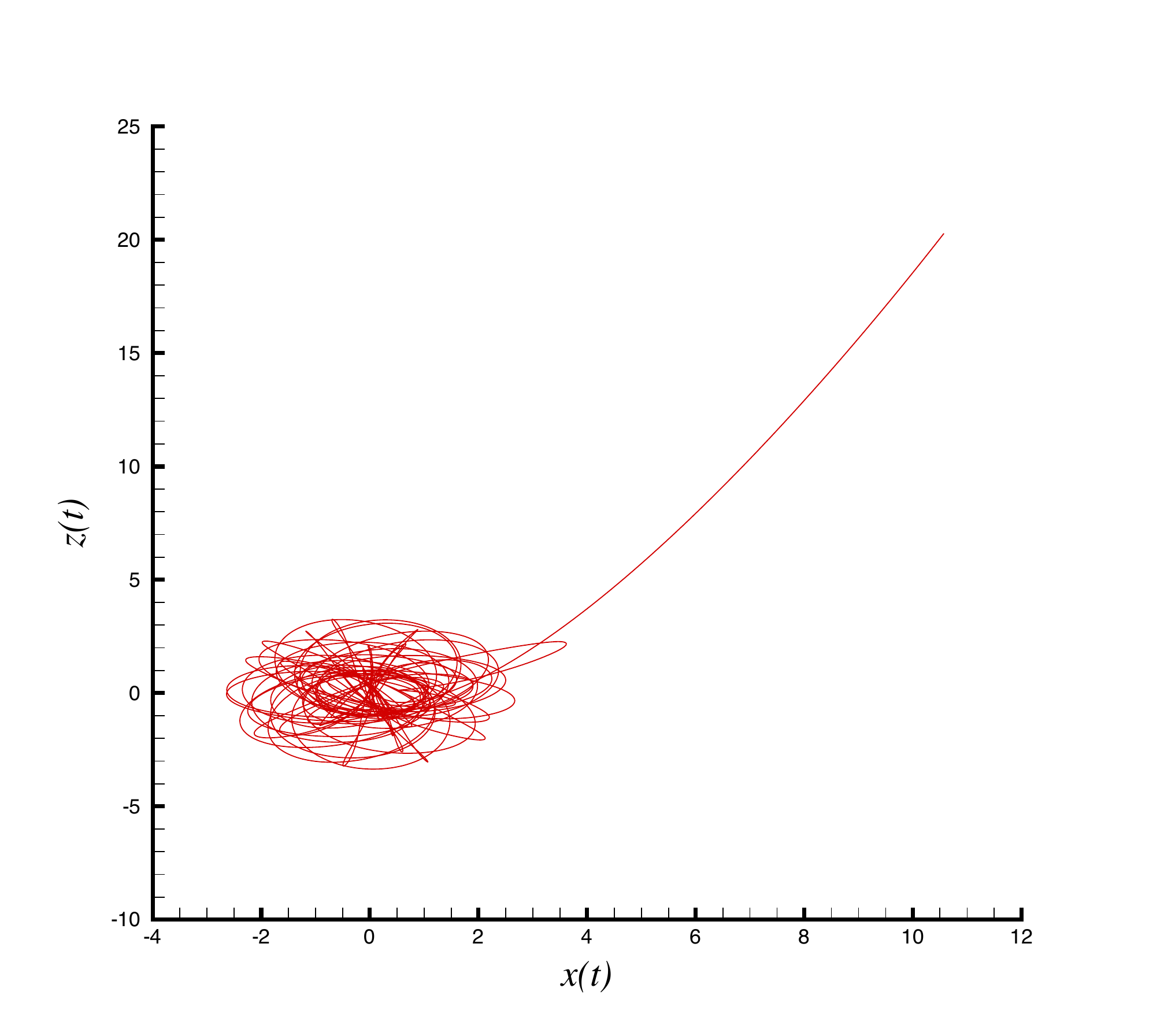}
\caption{ $x-y$ and $x-z$ of Body 1  ($0 \leq t \leq 1000$) in the case of $\delta = 10^{-60}$. }
\label{figure:body1dX-2D}

\centering
\includegraphics[scale=0.3]{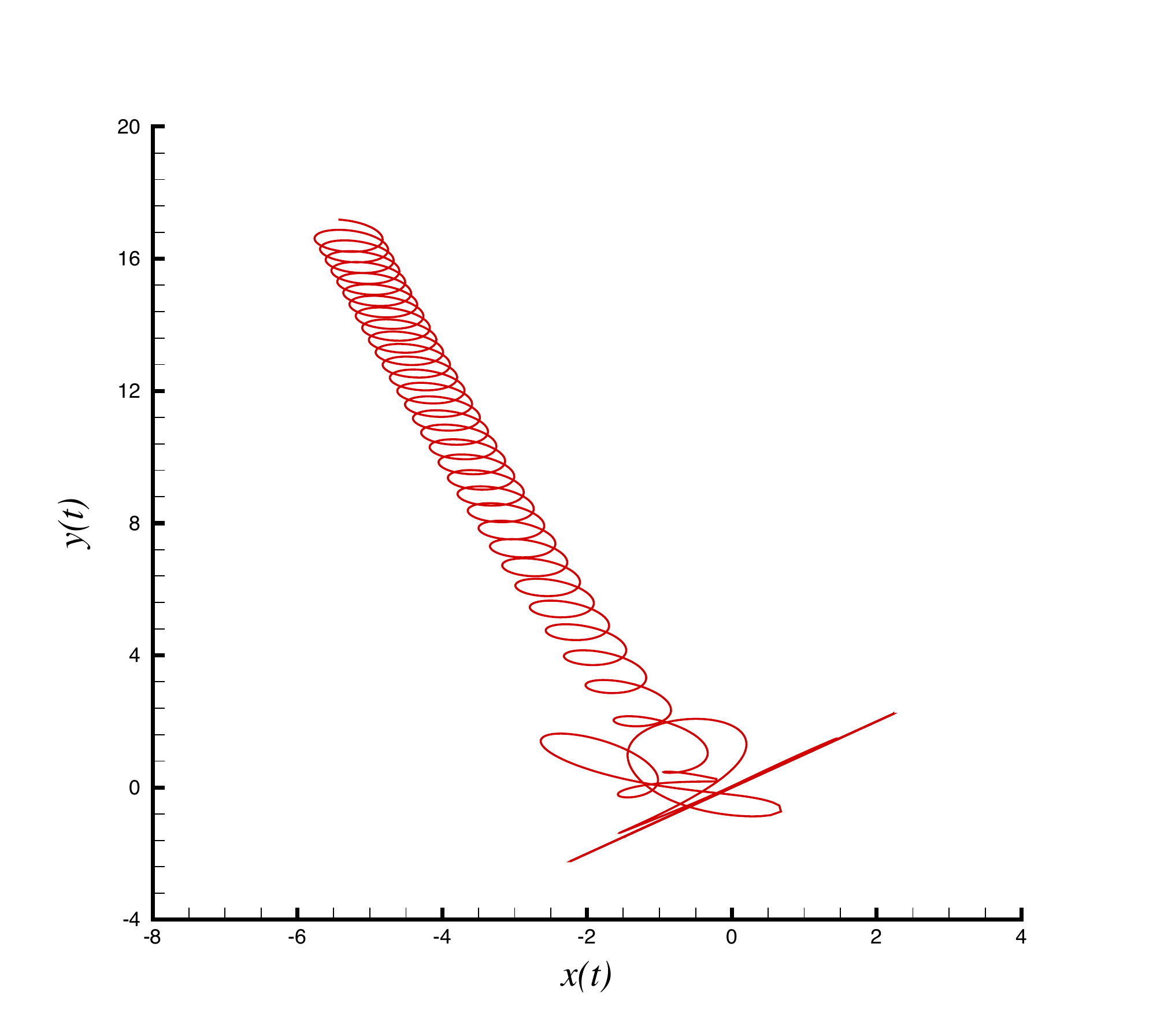}
\includegraphics[scale=0.3]{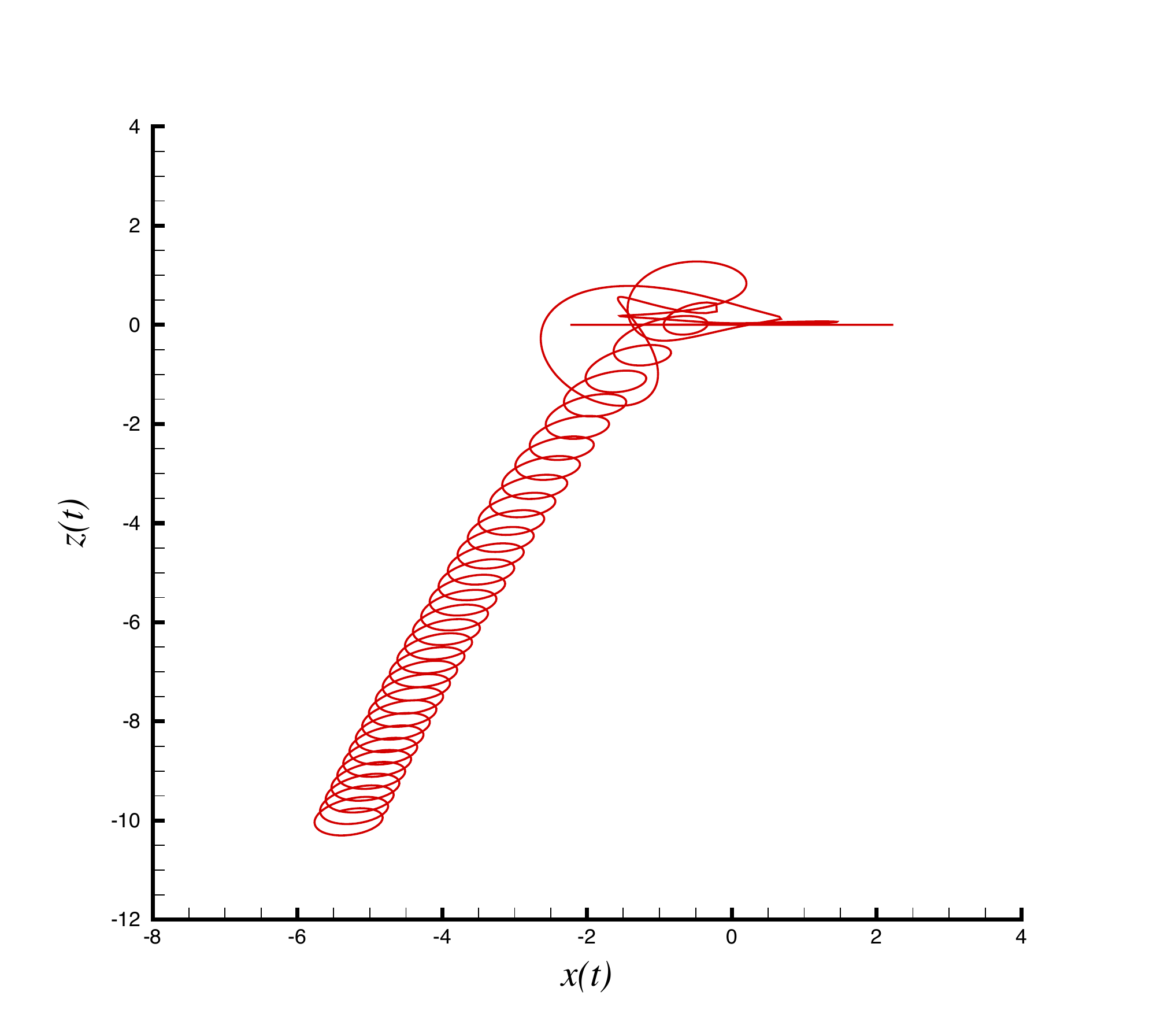}
\caption{ $x-y$ and $x-z$ of Body 2  ($0 \leq t \leq 1000$) in the case of $\delta = 10^{-60}$. }
\label{figure:body2dX-2D}

\centering
\includegraphics[scale=0.3]{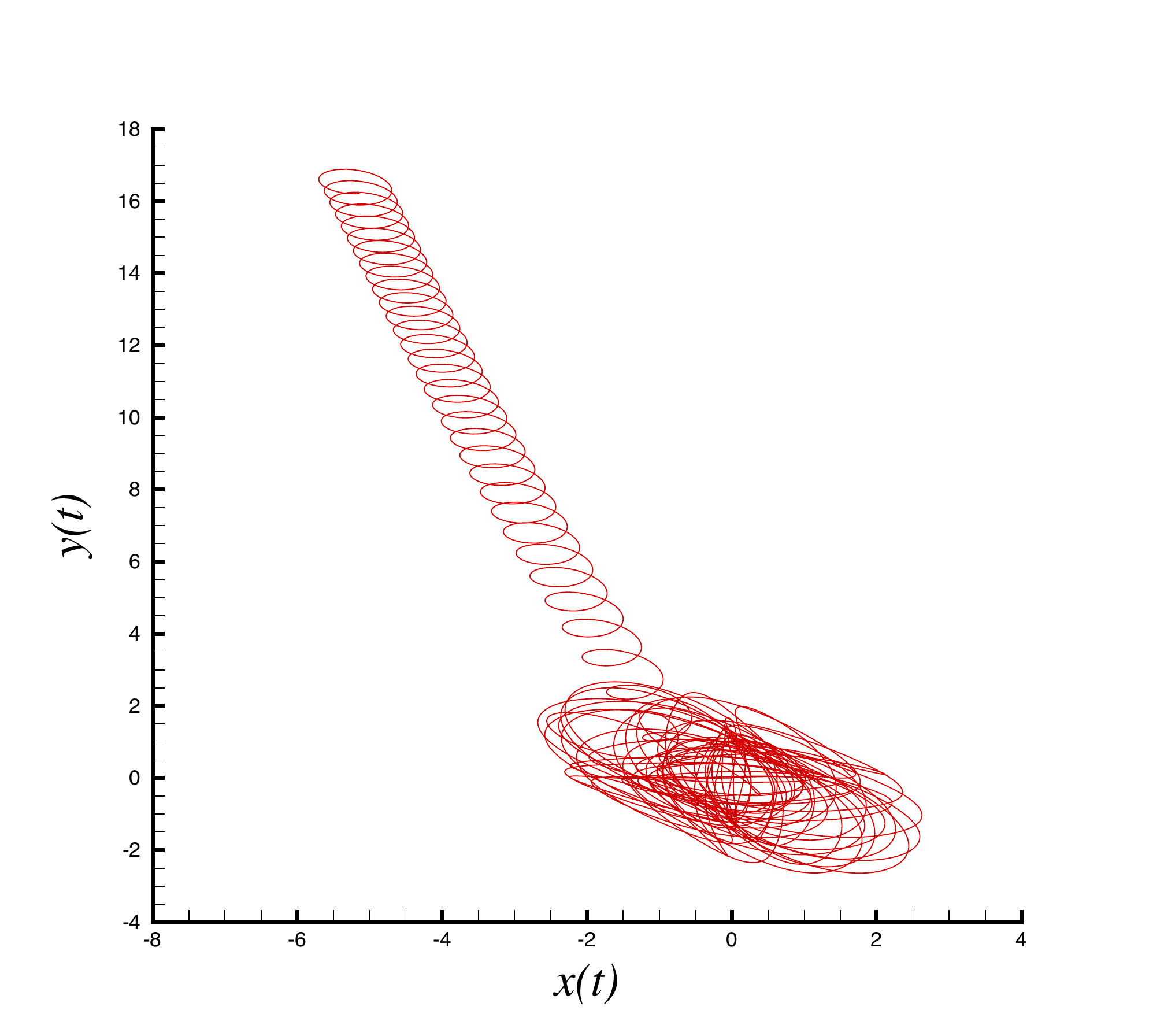}
\includegraphics[scale=0.3]{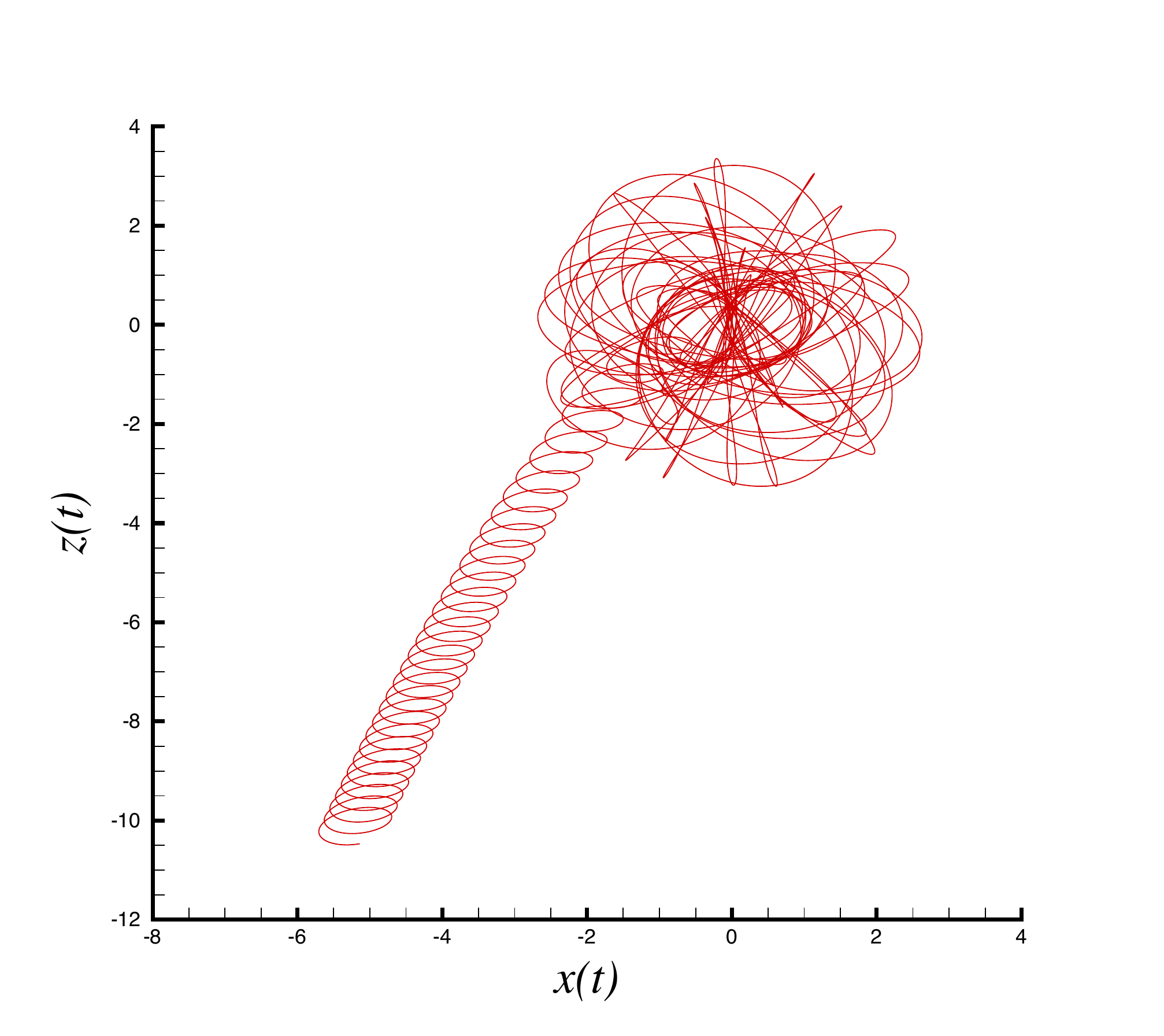}
\caption{ $x-y$ and $x-z$ of Body 3  ($0 \leq t \leq 1000$) in the case of $\delta = 10^{-60}$. }
\label{figure:body3dX-2D}
\end{figure}

\begin{figure}
\centering
\includegraphics[scale=0.3]{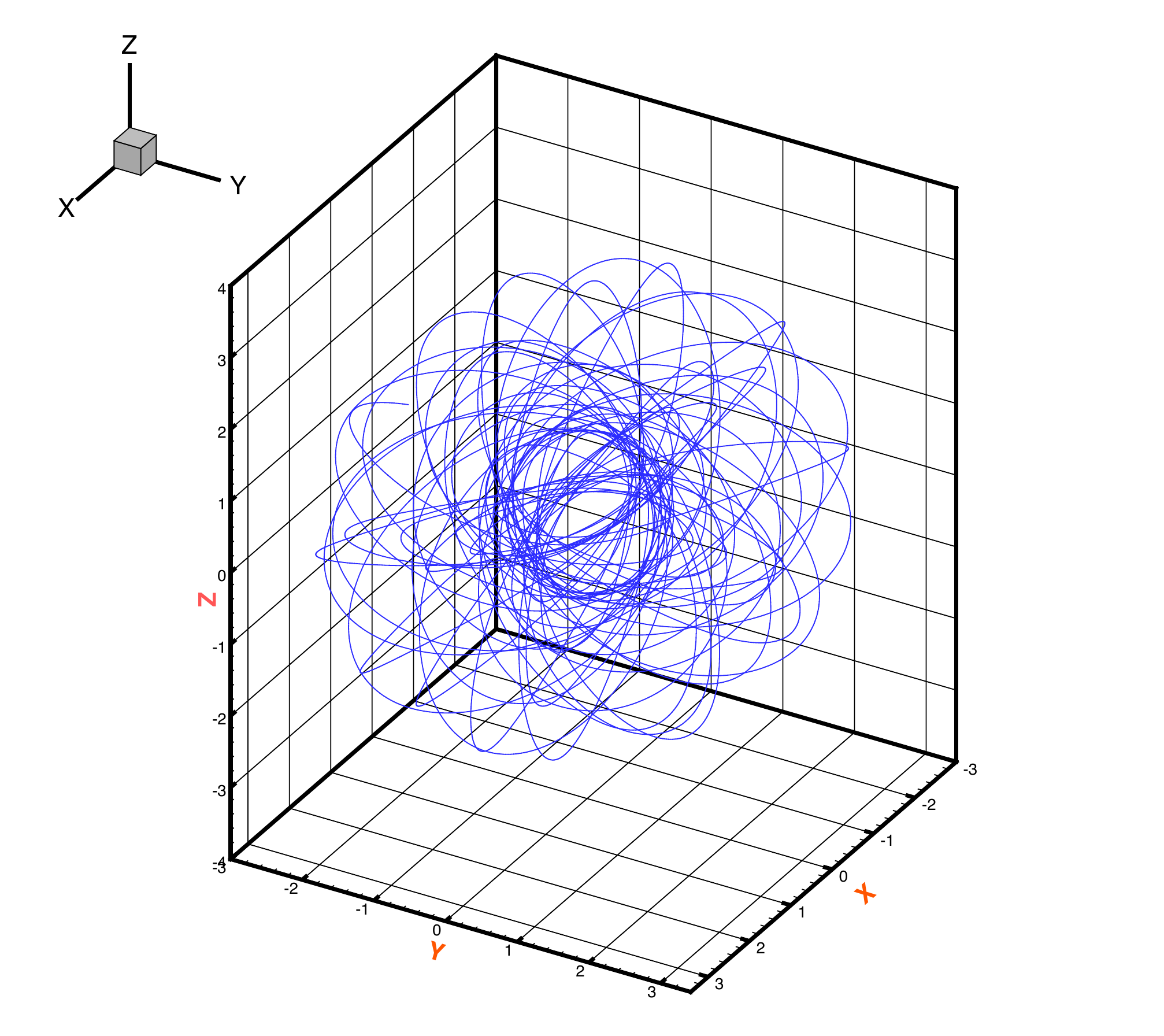}
\includegraphics[scale=0.3]{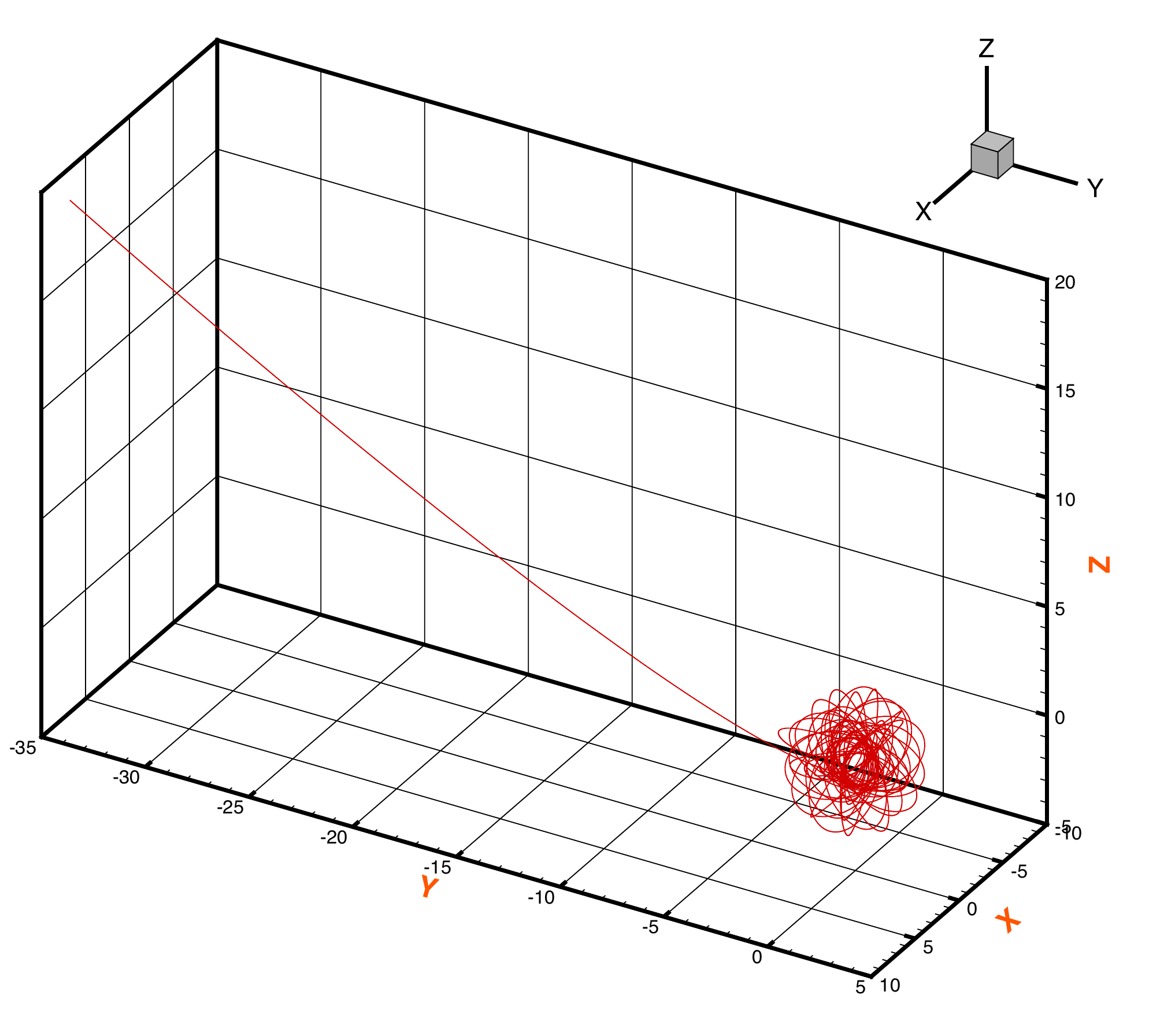}
\caption{  Orbit of Body 1 ($0 \leq t \leq 1000$).  Left: $\delta = 0$;  Right: $\delta = 10^{-60}$. }
\label{figure:body1-3D}

\centering
\includegraphics[scale=0.3]{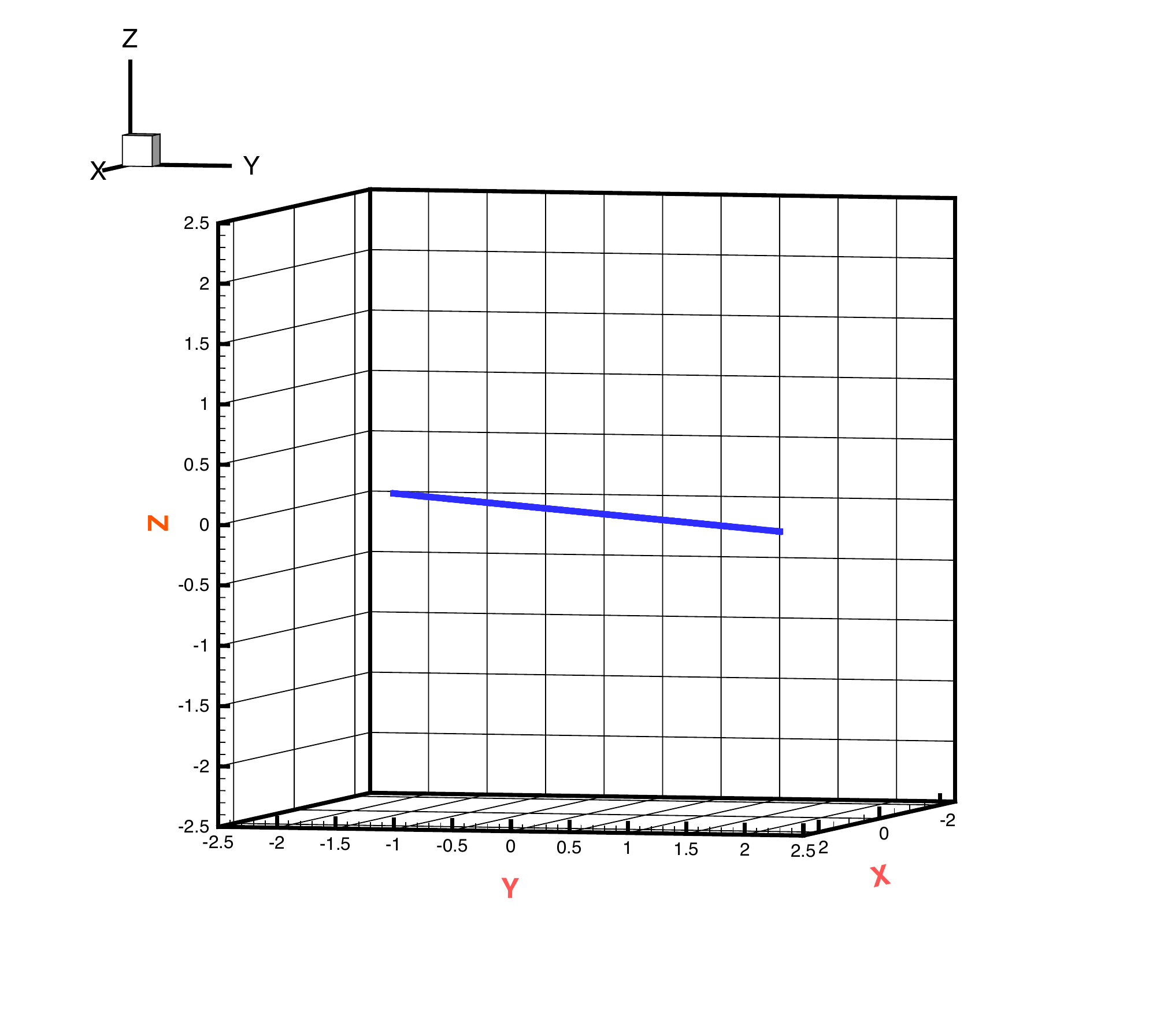}
\includegraphics[scale=0.3]{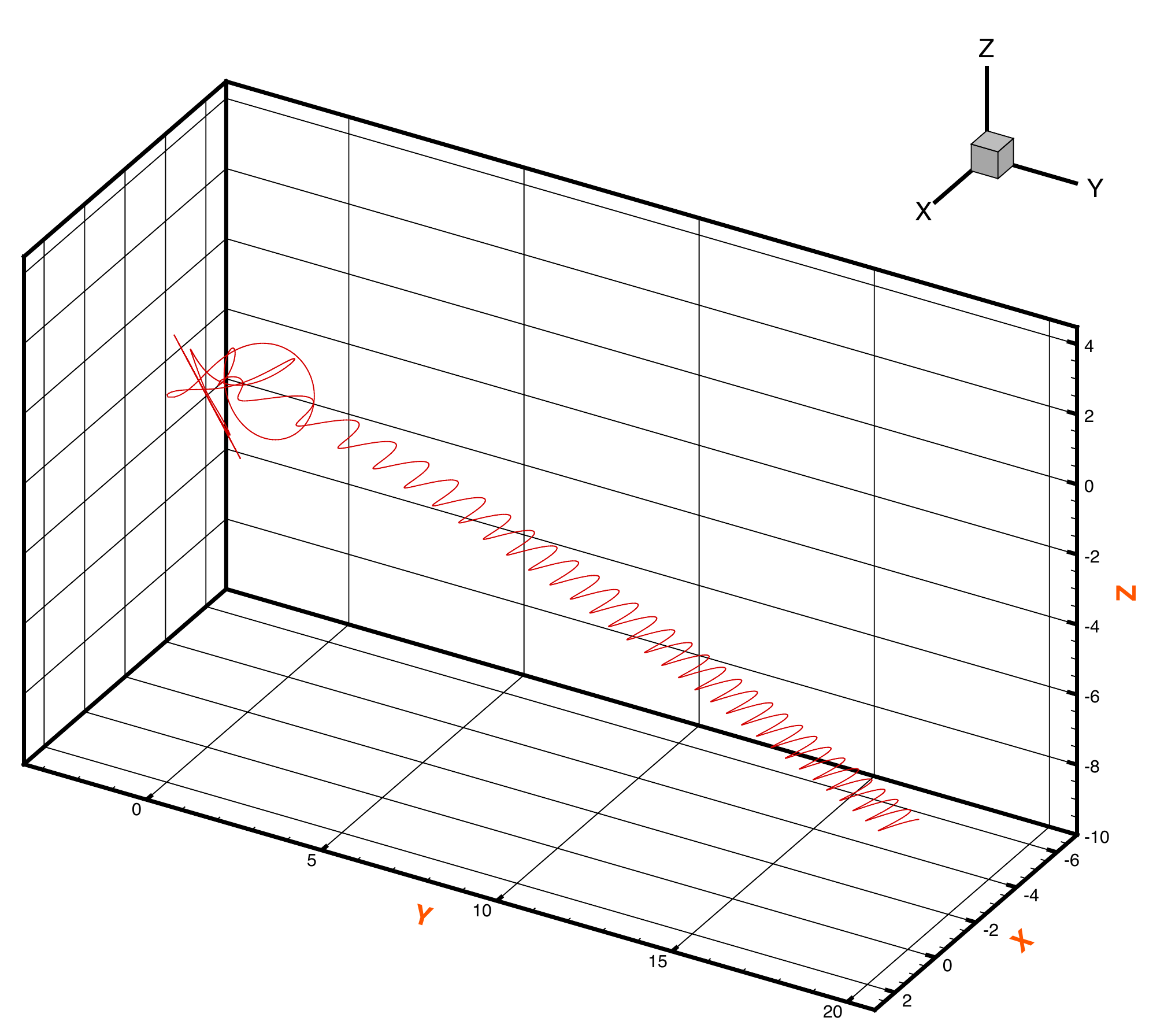}
\caption{  Orbit of Body 2  ($0 \leq t \leq 1000$).   Left: $\delta = 0$;  Right: $\delta = 10^{-60}$.  }
\label{figure:body2-3D}

\centering
\includegraphics[scale=0.3]{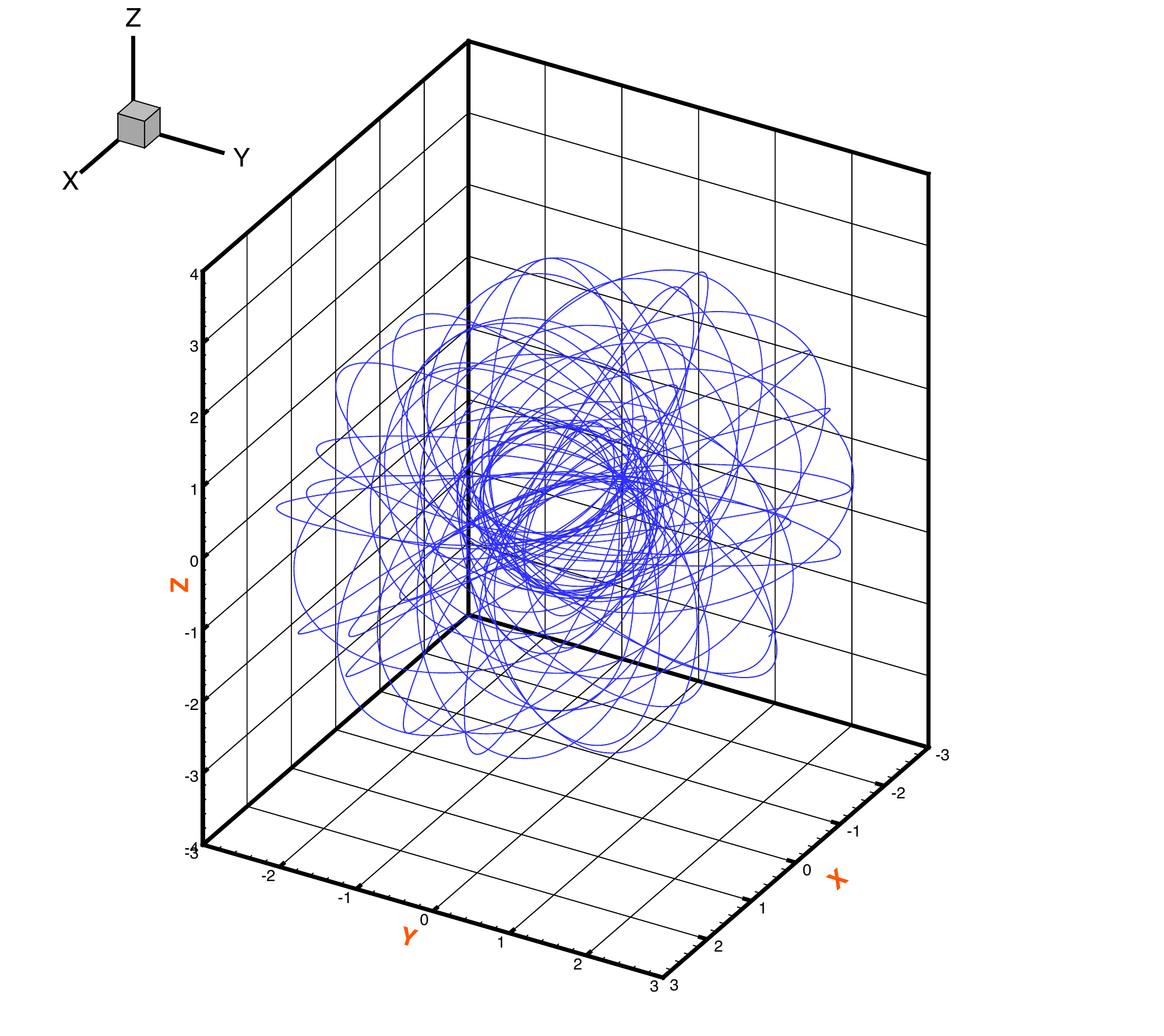}
\includegraphics[scale=0.3]{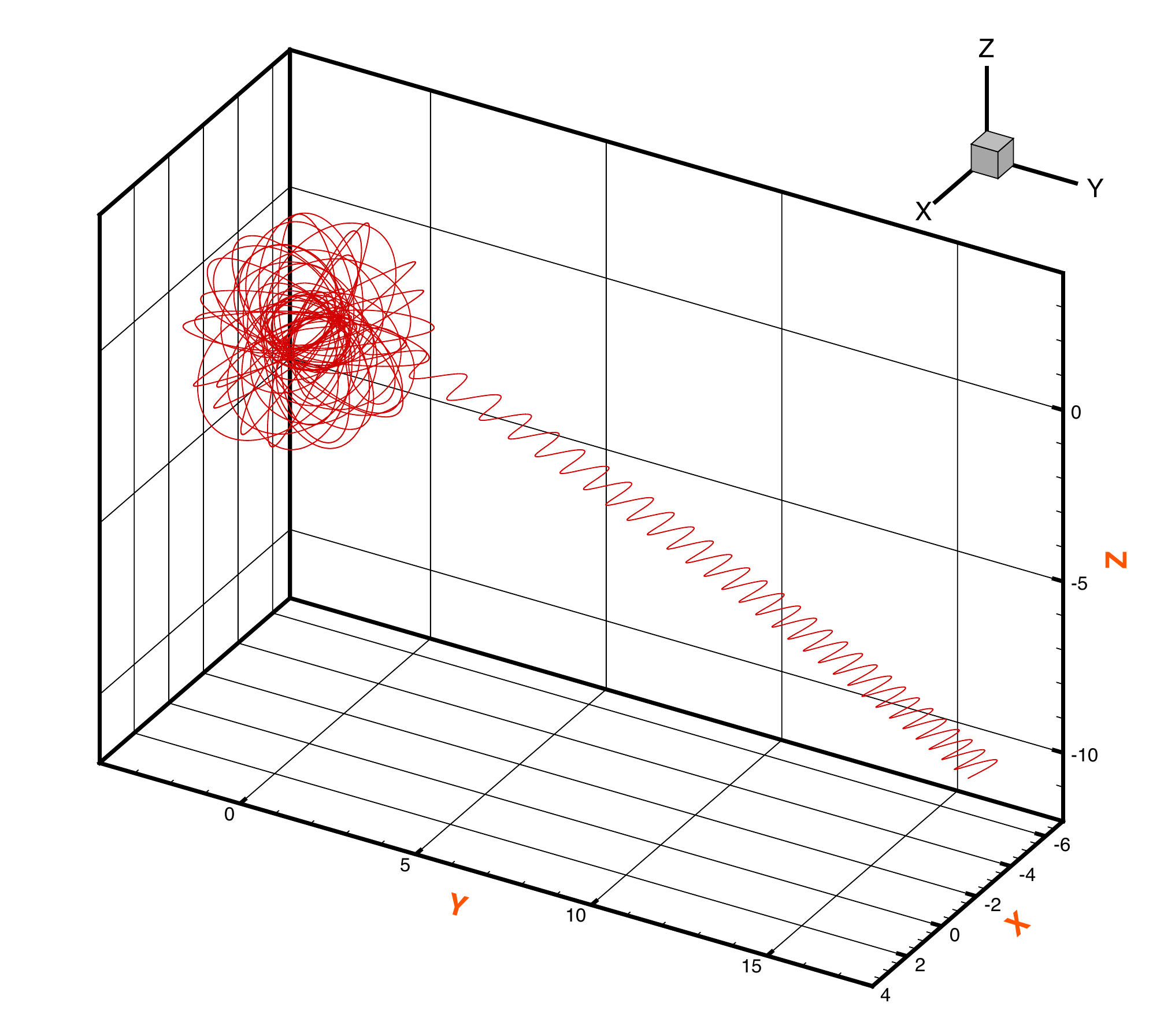}
\caption{  Orbit of Body 3  ($0 \leq t \leq 1000$).   Left: $\delta = 0$;  Right: $\delta = 10^{-60}$.  }
\label{figure:body3-3D}
\end{figure}

The initial  conditions when $\delta=10^{-60}$ have  a tiny difference \[  d {\bf r}_1 = 10^{-60} (1,0,0)\]  from those  when $\delta = 0$.  Thus,  it is reasonable to assume that the corresponding dynamic system is chaotic, too.  Similarly, the corresponding  orbits of the three bodies can be accurately  simulated  by means of the CNS.    It is found that, when $\delta=10^{-60}$,   the CNS results at $t = 1000 $ by means of $\Delta t = 10^{-2}$, $N=300$  and $M$ = 16, 24, 30, 40, 50, 60 , 70, 80, 100  agree  each other   in the accuracy of  1, 3, 5, 9, 11, 14, 17, 19 and 27  significance digits,  respectively.    Approximately, $n_s$ (the number of significance digits) is linearly  proportional to $M$ (the order of Taylor expansion), say, $n_s \approx 0.2885 M - 3.4684$, as shown in Fig.~\ref{figure:accuracy:delta-60}.   According to this formula, in order to have the CNS results (at $t=1000$) in the precision of 81 significance digits by means of  $\Delta t=10^{-2}$,  the   300th-order of Taylor expansion, i.e. $M=300$, must be used.  
But, this needs much more CPU time.   To  confirm the correction of these CNS results,  we further  use  the smaller time step $\Delta t=10^{-3}$.   It is found that, 
when $\delta=10^{-60}$ , the results at $t = 1000 $  by means of the CNS using $\Delta t = 10^{-3}$, $N=300$ and  $M$ = 8, 16, 24, 30,  40, 50 agree well  in the precision of  8,  21, 33, 43, 59, 72 significance digits,  respectively.    Approximately, $n_s$, the number of significance digits of the corresponding results at $t=1000$, is linearly  proportional to $M$ (the order of Taylor expansion), say, $n_s \approx1.5386 M - 3.7472$, as shown in Fig.~\ref{figure:accuracy:delta-60}.   
For example,  the  position of Body 1 at $t = 1000 $ given by the 50th-order Taylor expansion and data in 300-digit precision with $\Delta t=10^{-3}$ reads
\begin{eqnarray}
x_{1,1} &=& + 10.57189 91771 62684 86053 99651 18023 33873 55185 816 \nonumber \\ 
&& 19 03577 63318 20966  52436 81014 64,  \label{x[1,1]-delta-60}\\
x_{2,1} &=& -33.39568 60196 58214 70317 81512 36176 86024 07559 680 \nonumber \\ 
&& 29 19927 61867 10038 14287 39294 80, \\
x_{3,1} &=& +  20.28455 27396 82192 29521 36869 21793 88441 04404 153 \nonumber \\
&& 94 34152 86710 55848 80509 72622 14,\label{x[3,1]-delta-60}
\end{eqnarray}
which are in the precision of 72 significance digits.    Note  that  the  positions of the  three bodies   at  $t=1000$  given by $\Delta t=10^{-2}$ and $M=100$ agree well (in precision of 27 significance digits) with those by $\Delta t=10^{-3}$ and  $M=50$.    In addition, the momentum conservation (\ref{conversation:r}) is satisfied in the level of $10^{-293}$.    Thus,  our  CNS  results  in the case of  $\delta=10^{-60}$  are  reliable  in  the  interval $0\leq t \leq 1000$ as well.

The orbits of the three bodies in the case of $\delta=10^{-60}$ are as shown in Figs.~\ref{figure:body1dX-2D} to \ref{figure:body3dX-2D}.   It is found that, in the time interval $0 \leq t \leq 800$,  the orbits of the three bodies are not obviously different from those in the case of $\delta = 0$, say, Body~2 oscillates along the same line on $z=0$, Body~1 and Body~3 are chaotic with the same symmetry about the regular orbit of Body~2.   However,  the obvious difference of  orbits  appears when $t\geq 810$: Body 2 departs from the oscillations along the line on $z=0$ and  escapes (together with Body~3) along a complicated three-dimensional  orbit.   Besides, Body~1 and  Body~3  escape  in  the opposite  direction without any symmetry.   As shown in Figs.~\ref{figure:body1-3D} to \ref{figure:body3-3D},  Body~2 and Body~3 go far and far away from Body~1 and thus  become a binary-body system.   Thus, it is very interesting that,  the tiny difference $d {\bf r}_1 =10^{-60}(1,0,0)$ of the initial conditions finally disrupts  not only the  elegant  symmetry  of  the orbits but also even the three-body system itself!   

\begin{figure}
\centering
\includegraphics[scale=0.3]{body1-dX-60.pdf}
\includegraphics[scale=0.3]{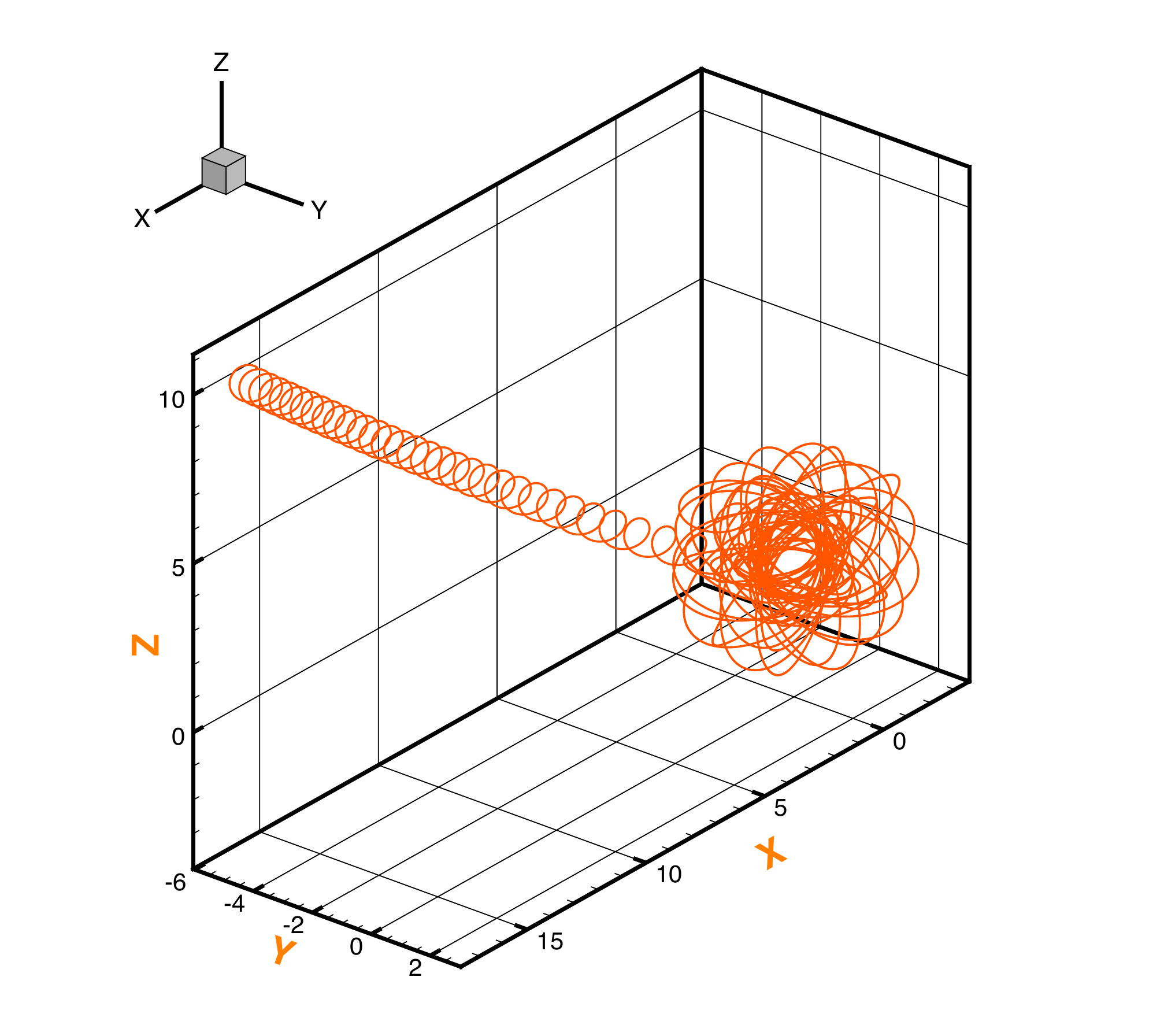}
\caption{  Orbit of Body 1 ($0 \leq t \leq 1000$).  Left: $\delta = +10^{-60}$;  Right: $\delta = -10^{-60}$. }
\label{figure:body1-3D-negative}

\centering
\includegraphics[scale=0.3]{body2-dX-60.pdf}
\includegraphics[scale=0.3]{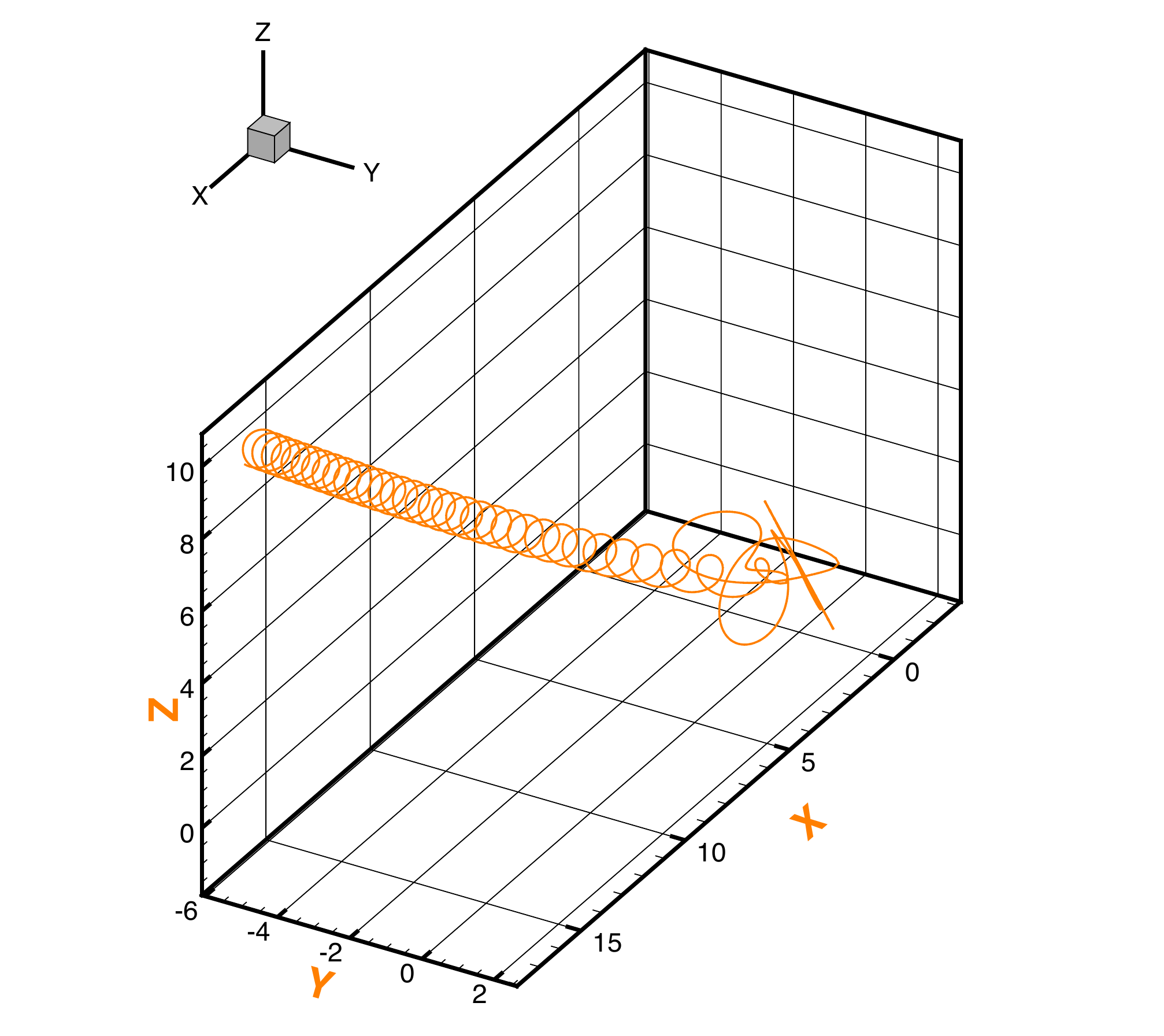}
\caption{  Orbit of Body 2  ($0 \leq t \leq 1000$).   Left: $\delta = +10^{-60}$;  Right: $\delta = -10^{-60}$.  }
\label{figure:body2-3D-negative}

\centering
\includegraphics[scale=0.3]{body3-dX-60.pdf}
\includegraphics[scale=0.3]{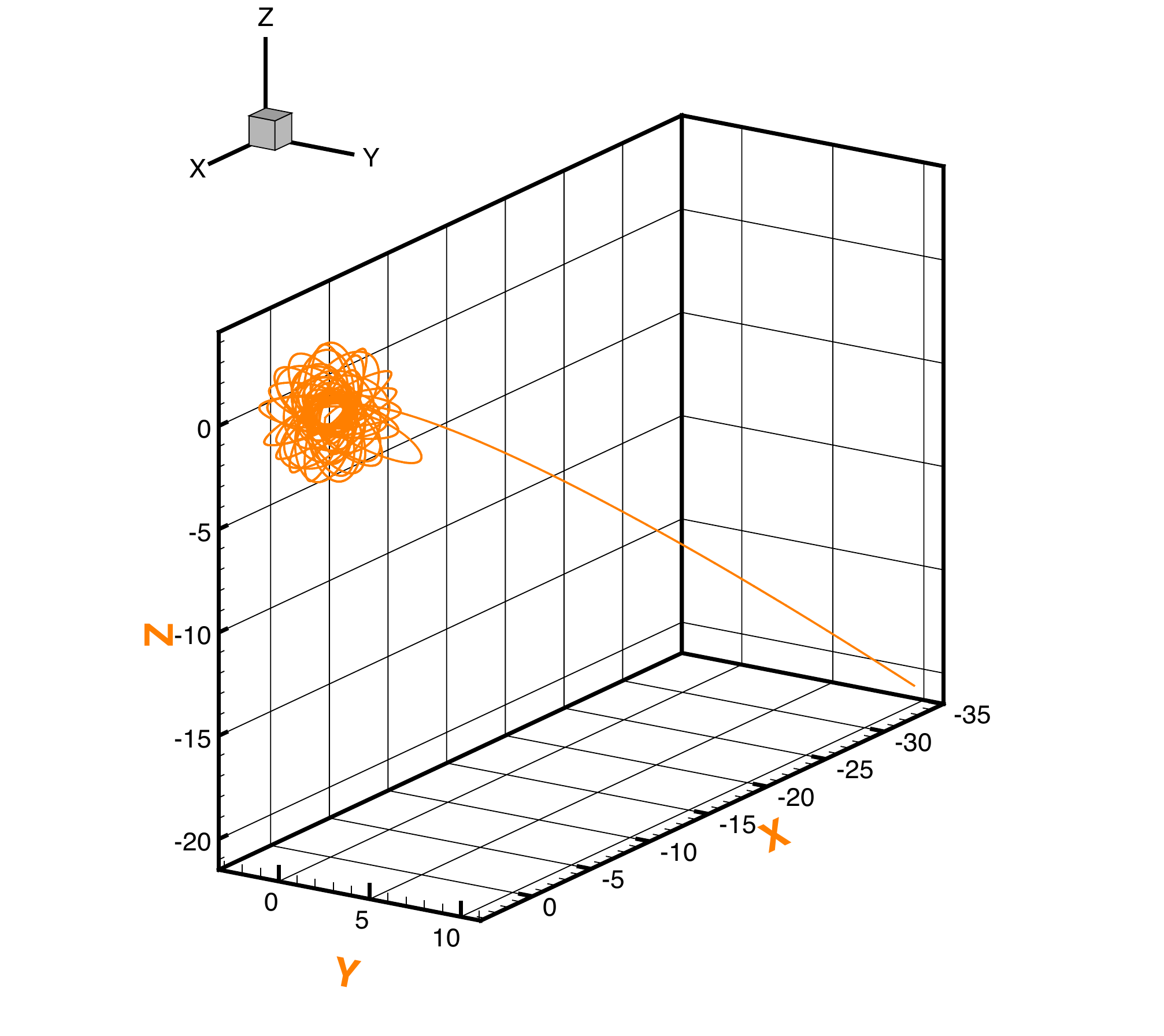}
\caption{  Orbit of Body 3  ($0 \leq t \leq 1000$).   Left: $\delta = +10^{-60}$;  Right: $\delta = -10^{-60}$.  }
\label{figure:body3-3D-negative}
\end{figure}

Similarly,  in the case of $\delta=-10^{-60}$,  we gain the reliable orbits of the three bodies by means of the CNS with $\Delta t =10^{-3}$, $N=300$ and $M=20$ (i.e. the 20th-order Taylor expansion).   As shown in Figs.~\ref{figure:body1-3D-negative} to \ref{figure:body3-3D-negative},  the tiny difference in the initial position disrupts not only the elegant symmetry of the orbits but also the three-body system itself as well:  when $t>810$,  Body~2 departs from its oscillation along the line on $z=0$ and escapes (but together with Body~1) in a complicated three-dimensional orbit, while Body~1 and Body~3  escape  in the opposite  direction   without  any  symmetry.    Note that,  in the case of $\delta=-10^{-60}$,  Body~1 and Body~2   go  together  far and far away  from Body~3  to become a binary system.  However, in the case of $\delta=+10^{-60}$, Body~2 and Body~3 escape together to become a binary system!  This is very interesting.   Thus,  the orbits of the three-body system when  $\delta = 0$,  $\delta = +10^{-60}$ and $\delta = -10^{-60}$ are completely different.

From the {\em mathematical} viewpoints,  the above results are not surprising at all and thus there exist nothing new:   since the three-body system is chaotic (as pointed out by  Sprott \cite{Sprott2010}),  the  results  are certainly  very  sensitive to the initial conditions.  However,  the difference of the three initial positions is so small that they can be regarded as the {\bf same} in physics!  In other words, from {\em physical} viewpoints,  such a small difference in space has no physical meanings at all, and thus  the three initial conditions (when $\delta=0$ and $\delta = \pm 10^{-60}$, respectively) are the {\bf same} in physics.  This is mainly because position of any a body   {\em inherently} contains  the  micro-level  uncertainty so that  the three-body system is {\em not} deterministic, as  explained below.

It is well known  that the microscopic  phenomenon are essentially uncertain/random.    Let us  first consider some typical length scales of microscopic  phenomenon which are widely used in modern physics.     For example, 
Bohr radius 
\[ r = \frac{\hbar^2}{m_e \; e^2} \approx 5.2917720859(36) \times 10^{-11} \;\; \mbox{(m)}\]
 is the approximate size of a hydrogen atom, where $\hbar$ is a reduced Planck's constant, $m_e$ is the electron mass, and $e$ is the elementary charge, respectively.   Besides,  the so-called  Planck length 
\begin{equation}
l_P = \sqrt{\frac{\hbar \; G}{c^3}} \approx 1.616252(81) \times 10^{-35} \;\;\; \mbox{(m)}
\end{equation}
is the length scale at which quantum mechanics, gravity and relativity all interact very strongly,  where  $c$ is the speed of light in a vacuum and $G$ is the gravitational constant.   According to the string theory \cite{Polchinshi1998}, the Planck length is the order of magnitude of oscillating strings that form elementary particles,  and  {\em shorter length do not make physical senses}.    Especially,  in some forms of quantum gravity, it becomes  {\em impossible}  to  {\em determine} the difference between two locations less than one Planck length apart.    Therefore,  in the  level  of the Planck length,  position of a body is  {\em inherently} uncertain.    This kind of microscopic  physical  uncertainty is {\em inherent} and has nothing to do with  Heisenberg uncertainty principle \cite{Heisenberg1927} and the ability of human being as well.

In addition,  according to de Broglie \cite{Broglie1924},  any  a  body has the so-called  wave-particle duality.   The de Broglie's wave of a body has non-zero amplitude.  Thus,  position of a body is uncertain:  it could be almost {\em anywhere}   along de Broglie's wave packet.   Thus,  according to  the de Broglie's wave-particle duality,  position of a star/planet is {\em inherent} uncertain, too.   Therefore,  it is  reasonable   to  assume that, from the physical viewpoint,   the micro-level inherent fluctuation of position of a body  shorter than the  Planck length $l_p$ is essentially  uncertain and/or  random.   
         
To make the  Planck length $l_p\approx 1.62 \times 10^{-35}$ (m)  dimensionless, we use the dimeter of Milky Way Galaxy as the characteristic length,  say,  $d_M  \approx 10^5$ light year $\approx 9 \times 10^{20}$ meter.  Obviously,  $l_p/d_M \approx 1.8 \times 10^{-56} $ is a rather small dimensionless number.   Thus, as mentioned above,  two (dimensionless) positions shorter than $10^{-56}$ do not make physical senses in many cases.  So,  it is reasonable to assume  that  the inherent  uncertainty of the dimensionless position  of a star/planet  is  in  the micro-level $10^{-60}$.   Therefore, the tiny difference $d {\bf r}_1 = \pm 10^{-60} (1,0,0)$ of the initial conditions is in the micro-level:  the difference is so small that all of these initial conditions can be regarded as the {\em same} in physics!   

Mathematically,   $10^{-60}$ is a tiny number,  which  is  much  smaller  than truncation and round-off errors of traditional numerical  approaches  based on  data of 16-digit precision.    So, it is impossible to investigate the influence and evaluation of this inherent micro-level uncertainty of initial conditions by means of the traditional numerical approaches.  However,  the micro-level uncertainty $10^{-60}$ is much larger than the truncation and round-off errors of the CNS  results  gained by means of  the high-order Taylor expansion and data in  300-digit precision with a reasonable time step $\Delta t$, as illustrated above.   So, the CNS provides us a convenient tool to study the transfer and evaluation of such kind of inherent micro-level uncertainty of initial conditions.       

The key point is that such an  inherent  micro-level  uncertainty in the initial conditions finally leads to the huge,  observable difference of orbits of the three bodies:  it  disrupts not only the elegant symmetry of the orbits  but also  the three-body system itself.   Note that,  Body~2 escapes  with Body~3 in the case of $\delta=+10^{-60}$,  but with Body~1 in the case of $\delta = -10^{-60}$, to become a binary-body system!   In nature,   such kind of inherent micro-level uncertainty exists for {\em each} body at {\em any} time $t\geq 0$.   So,  from the physical viewpoint,   the orbits of each body at large enough time are inherently unknown, i.e. random:  given the {\em same} initial condition (in the viewpoint of physics), the orbits of the three-body system under consideration  might be completely {\em different}.   For example,   the three-body system might either have the elegant symmetry, or disrupt as the different  binary-body systems,  as shown in Figs.~\ref{figure:body1-3D} to \ref{figure:body3-3D}, and Figs.~\ref{figure:body1-3D-negative} to \ref{figure:body3-3D-negative}, respectively.

Note that, from {\em mathematical} viewpoint, we can accurately simulate the orbits of three bodies in the interval $0 \leq t \leq 1000$.   However,  due to the inherent position uncertainty and the SDIC of chaos,   orbits of  the  three bodies   are random when $t>810$, since the inherent position uncertainty transfers into macroscopic randomness.    Thus, from {\em physical}  viewpoint, there exists  the maximum predictable time $T^p_{max}$,  beyond which the orbits of the three bodies are inherently random and thus can {\em not} be predictable in essence.   Note that  $T^p_{max}$ is determined by the {\em inherent}  position  uncertainty of the three bodies, which has {\em nothing} to do with the ability of human being.   Therefore, long term ``prediction'' of chaotic dynamic system of the three bodies is {\em mathematically } possible,  but has no  {\em physical}  meanings!

\begin{figure}
\centering
\includegraphics[scale=0.3]{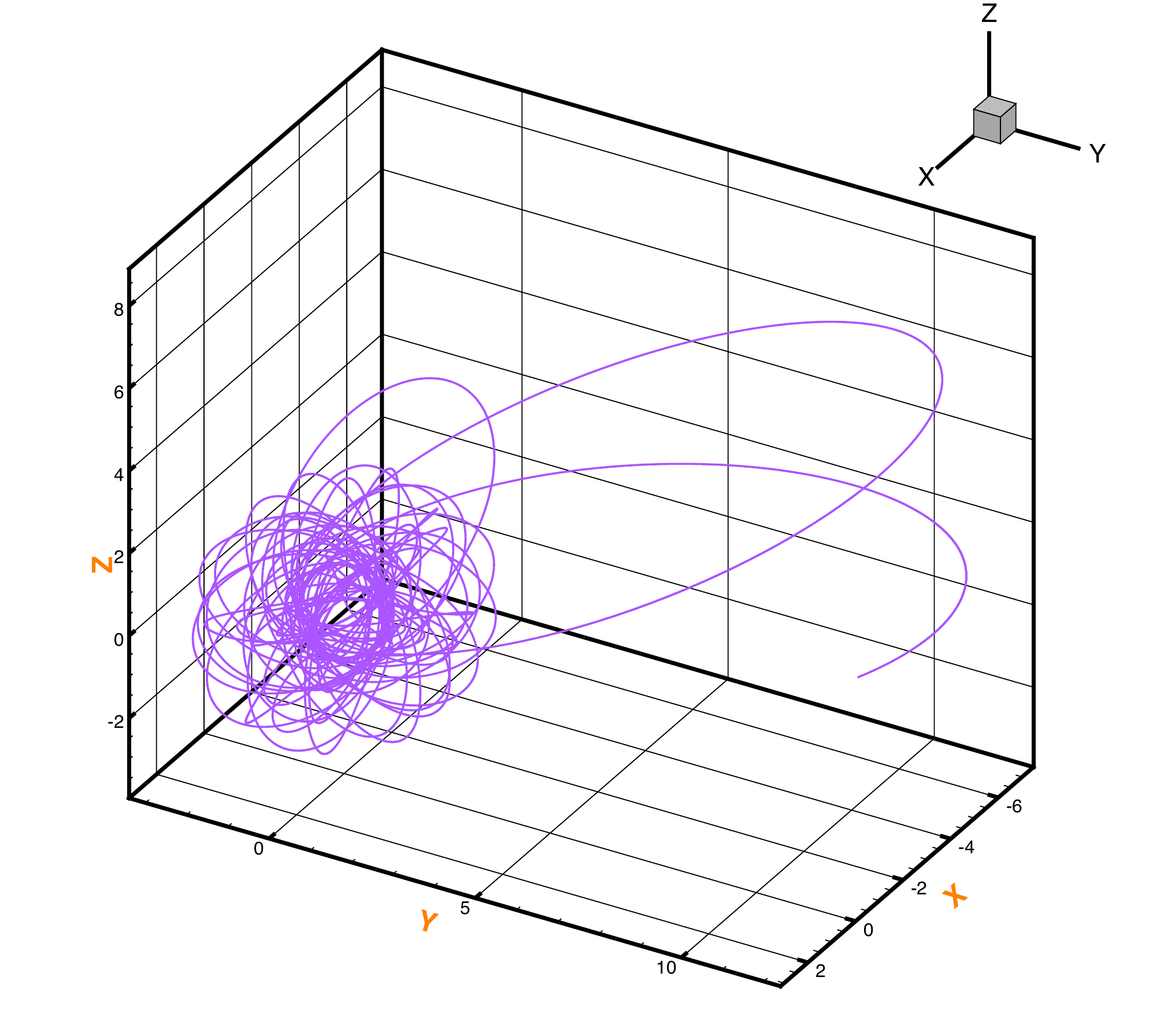}
\caption{  Orbit of Body 1 ($0\leq t\leq 1200$) when $d {\bf r}_1 = 10^{-60}\; \; (1,1,1)$. }
\label{figure:body1-3D-XYZ}

\centering
\includegraphics[scale=0.3]{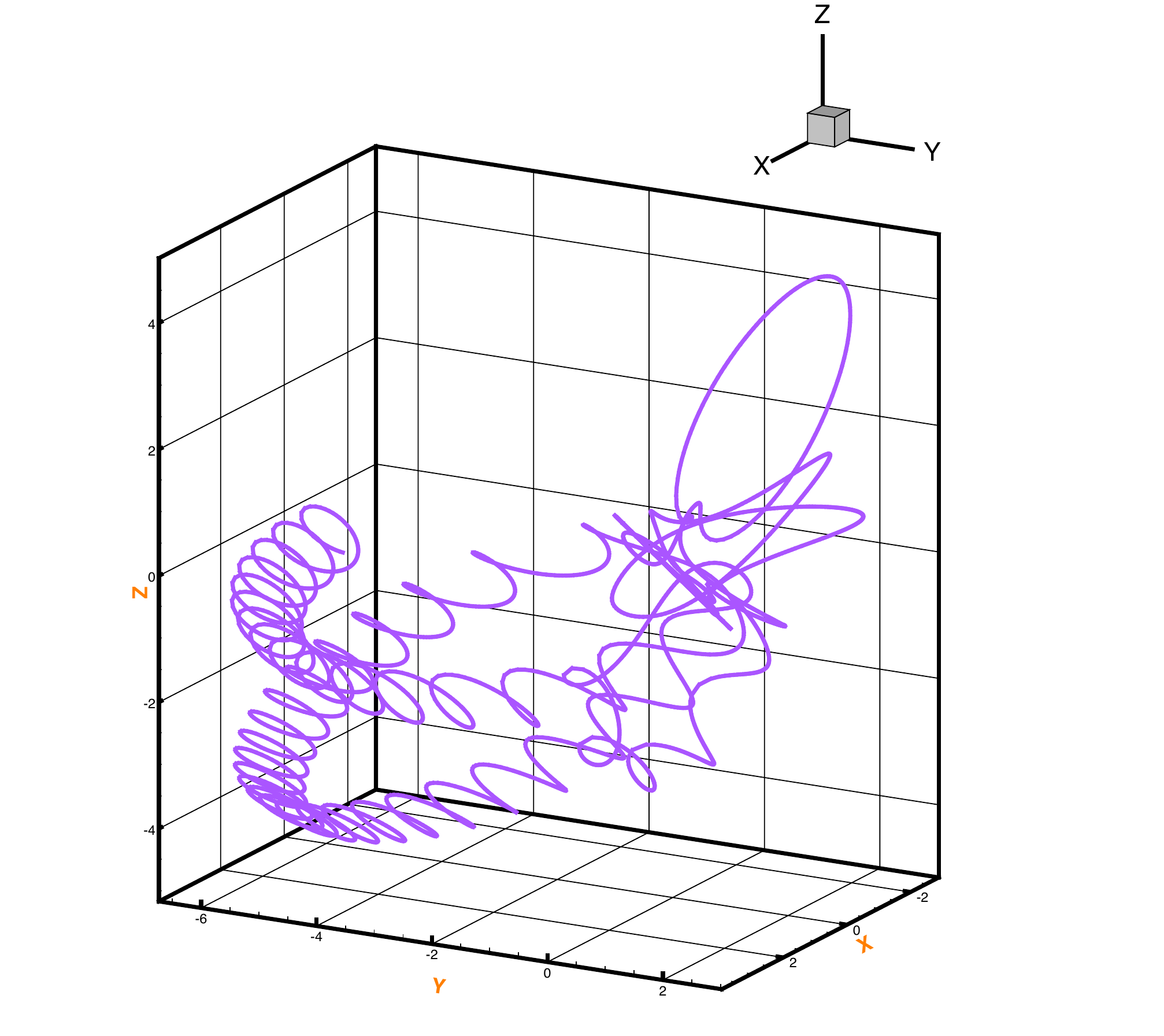}
\caption{  Orbit of Body 2 ($0\leq t\leq 1200$) when $d {\bf r}_1 = 10^{-60}\; \; (1,1,1)$. }
\label{figure:body2-3D-XYZ}

\centering
\includegraphics[scale=0.3]{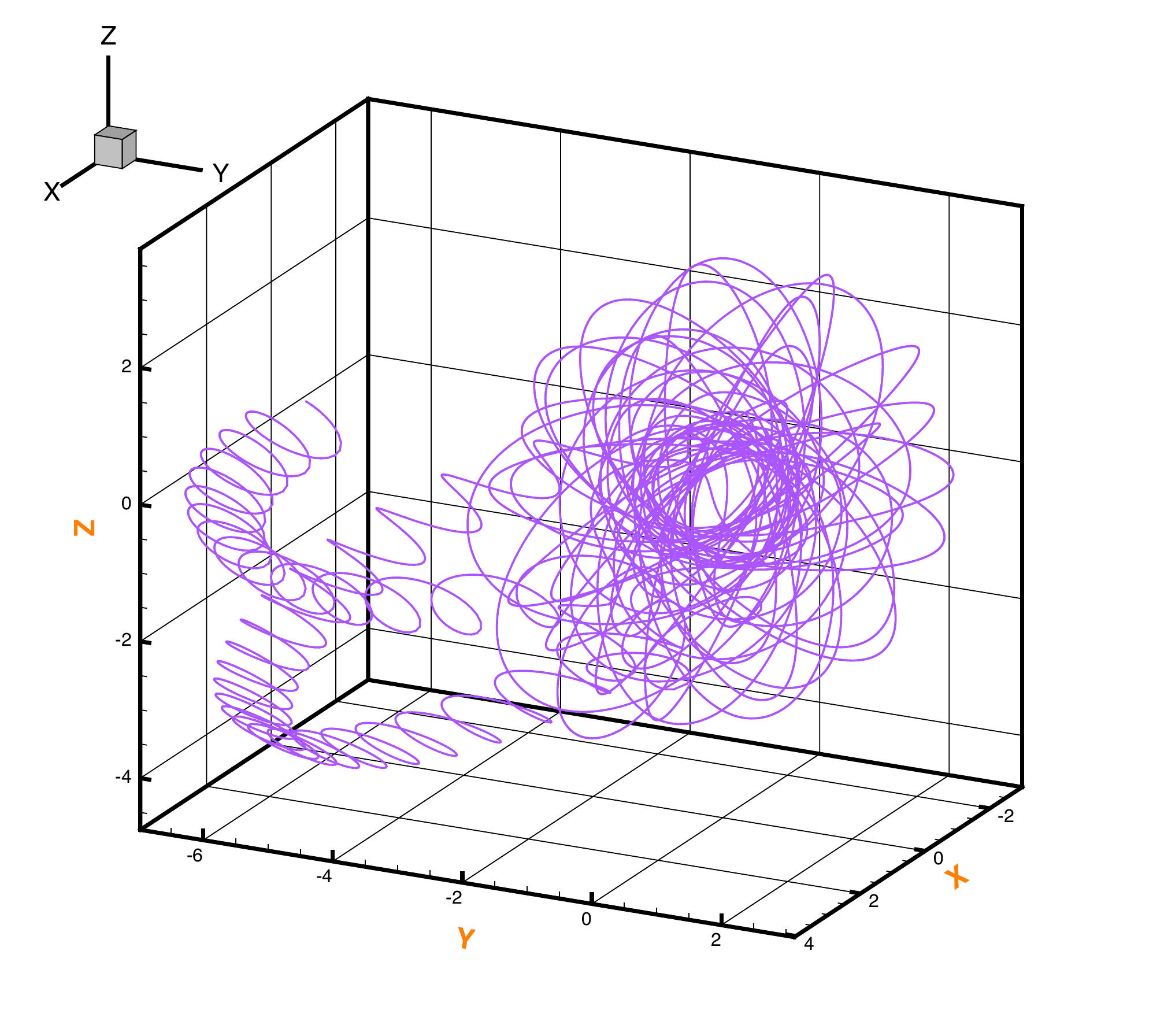}
\caption{  Orbit of Body 3 ($0\leq t\leq 1200$) when $d {\bf r}_1 = 10^{-60}\; \; (1,1,1)$. }
\label{figure:body3-3D-XYZ}
\end{figure}

Finally, to confirm our above conclusions,  we further consider  such a special case with the micro-level uncertainty of the initial position $d {\bf r}_1 = 10^{-60} (1, 1, 1)$, i.e. 
\[   {\bf r}_1 = (0,0,-1) + 10^{-60} (1, 1, 1).  \]
It is so tiny that, from the physical viewpoint mentioned above,  the initial positions can be regarded as the same as those of the above-mentioned three cases.  However,  the corresponding  orbits of the three bodies (in the time interval $0\leq t \leq 1200$) obtained by means of the CNS with $\Delta t =10^{-3}, N= 300$ and the 30th-order Taylor expansion ($M=30$) are quite different from those of the three cases: Body~2 first oscillates along a line on $z=0$ but  departs  from the regular orbit   for large $t$ to move along a complicated three-dimensional orbits, while Body~1 and Body~3 first move with the symmetry but lose it for large $t$, as shown in Figs.~\ref{figure:body1-3D-XYZ} to \ref{figure:body3-3D-XYZ}.  However, it is not clear whether any one of  them might escape or not,  i.e.  the fate of the three-body system is unknown.    Since  such  kind  of  micro-level uncertainty of position is inherent and unknown,  given the {\em same} (from the physical viewpoint) initial positions of the three bodies, the orbits of the three-body system at large enough time is completely unknown.   So, it has no physical meanings to talk about the accurate long-term prediction of orbits of the three-body system, because the orbits at  large time  (such as $t \geq 1000$) is  {\em inherently}  unknown/random  and  thus  should  be described by  probability.     This is quite similar to the motion of electron in an atom.    It should be emphasized that,  such kind of transfer from  the inherent micro-level uncertainty to macroscopic randomness  is  essentially  due to the SDIC of chaos, but has nothing to do with the ability of human being and Heisenberg uncertainty principle \cite{Heisenberg1927}.     

All of these  reliable  CNS  results  indicate that, due to the SDIC of chaos, such kind of inherent micro-level uncertainty of a star/planet might transfer into macroscopic randomness.   This provides us an explanation  for the macroscopic randomness of the universe, say,  the inherent micro-level uncertainty might be an {\em origin} of the microscopic randomness, although it might be not the unique one.    This might enrich and deepen our understandings about  not only the three-body problem but also the chaos.           

 \section{Conclusions} 
 
The famous three-body problem is investigated by means of a numerical approach with negligible numerical noises in a long enough time interval,  namely the Clean Numerical Simulation (CNS).    From physical viewpoints,  position of any bodies contains inherent micro-level uncertainty.   The evaluations of such kind of inherent micro-level uncertainty are accurately  simulated  by means of the CNS.   Our  reliable, very accurate CNS results  indicate that the inherent micro-level uncertainty of position of a star/planet might transfer into macroscopic randomness.  Thus, the inherent micro-level uncertainty of a body might be an origin of  macroscopic randomness of the universe.  In addition, from physical viewpoints,  orbits of some three-body systems at large time are  inherently random, and thus it has no physical meanings to talk about the accurate long-term prediction of  the  chaotic  orbits.   Note that such kind of  uncertainty and randomness has nothing to do with the ability of human being  and Heisenberg uncertainty principle \cite{Heisenberg1927}.

In this article, we introduce a new concept: the maximum predictable time $T^p_{max}$ of chaotic dynamic systems with physical meanings.   For the  chaotic motions of the  three body problem considered in this article,  there exists the so-called maximum predictable time $T^p_{max}\approx 810$,  beyond which the motion of three bodies  is  sensitive  to  the  micro-level  {\em inherent} uncertainty of position and thus becomes  {\em inherently} random  in physics.   The so-called maximum predictable time of chaotic three bodies  is determined by the {\em inherent} uncertainty of position and has {\em nothing} to do with the ability of human being.       Note that,  from the mathematically viewpoint,    we can accurately calculate the orbits  by means of the CNS in the interval $0\leq t \leq 1000$.    Thus,   considering the physical inherent uncertainty of position,  long term ``prediction'' of chaotic motion of  the three bodies  is  {\em mathematically} possible but has no {\em physical} meanings.  

In summary, our rather accurate computations based on the CNS about the famous three-body problem illustrate that,  the inherent  micro-level uncertainty of positions of starts/planets might be one origin of  macroscopic randomness of the universe.    This might enrich our knowledge and deepen our understandings about not only the three-body problem but also chaos.   Indeed, the reliable computations based on the CNS are helpful for us to understand the world better.  

Note that the computation ability of  human  being  plays an important role in the development of chaotic dynamic systems.   The finding of  SDIC of chaos  by Lorenz in 1963  is impossible without digit computer, although data used by Lorenz  in his  pioneering work is only in accuracy of 16-digits precision.   So, the CNS with negligible numerical noises provides us a useful tool to understand chaos better.

Finally, as reported by Sussman and Jack \cite{Sussman1988, Sussman1992},  the motion of Pluto and even the solar system is chaotic  with a time scale in the range  of  3  to 30 million years.   Thus,  due to the SDIC of chaos and the micro-level inherent uncertainty of positions of planets,  the solar system is in essence random.  Note that such kind of randomness has nothing to do with the  ability of human:  the history of human being  is  indeed  too short, compared to the time scale  of such kind of macroscopic randomness.    The determinism  is  only a  concept  of human being:  considering the  much shorter time-scale of the human being, one can still regard the solar system to be deterministic, even although it is  random  in essence.

 \section*{Acknowledgement} This work is partly supported by  the State Key Lab of Ocean Engineering (Approval No. GKZD010056-6) and 
 the National Natural Science Foundation of China.
 

\end{document}